\documentclass[aps, twocolumn, prb,preprintnumbers,superscriptaddress]{revtex4-2}

\usepackage{amsmath,amsfonts,amssymb,graphics,graphicx,epsfig,color,times,indentfirst,layout,hyperref,subfigure,multirow,bm}
\usepackage{hyperref}
\usepackage{bbold}
\usepackage{bm}
\hypersetup{linktocpage,colorlinks=true,citecolor=red}
\usepackage{soul}

\usepackage{graphicx}
\usepackage{dcolumn}
\usepackage{bm}
\usepackage{amssymb}
\usepackage{amsmath}
\usepackage{xcolor}
\usepackage{mathrsfs}
\usepackage{fixmath}
\hypersetup{colorlinks=true}
\usepackage[all]{hypcap} % let hyperlinks correctly point to figures rather than their captions; must be loaded after the hyperref package!

\usepackage[final]{changes}
\newcommand{\stkout}[1]{\ifmmode\text{\sout{\ensuremath{#1}}}\else\sout{#1}\fi}
\setdeletedmarkup{\stkout{#1}}

\newcommand{\nn}{\nonumber \\}

\newcommand{\mbd}{\mathbold}
\newcommand{\sia}{\textit{SI Appendix}}

\begin{document}

\title{Topological Kondo Superconductors}
%\title{Topological heavy fermion superconductor}

\author{Yung-Yeh Chang}
\affiliation{Physics Division, National Center for Theoretical Sciences, Hsinchu 30013, Taiwan Republic of China}
\affiliation{Department of Electrophysics, National Yang Ming Chiao Tung University, Hsinchu 30010, Taiwan Republic of China}
\author{Khoe Van Nguyen}
\affiliation{Department of Electrophysics, National Yang Ming Chiao Tung University, Hsinchu 30010, Taiwan Republic of China}
\author{Kuang-Lung Chen}
\affiliation{Department of Electrophysics, National Yang Ming Chiao Tung University, Hsinchu 30010, Taiwan Republic of China}
\author{Yen-Wen Lu}
\affiliation{Department of Physics and Astronomy, University of California, Riverside, California 92511, U.S.A.}
\author{Chung-Yu Mou}
\affiliation{Department of Physics, National Tsing Hua University, Hsinchu 30043, Taiwan Republic of China}
\author{Chung-Hou Chung}
\affiliation{Department of Electrophysics, National Yang Ming Chiao Tung University, Hsinchu 30010, Taiwan Republic of China}

\date{\today}

\begin{abstract}
%We theoretically propose a topological Kondo superconducting phase in a two-dimensional Kondo lattice.  Due to the odd parity of the localized $f$-orbitals, the Kondo hybridization between f-electrons and conduction electrons shows an odd-parity momentum dependent form factor.  Consequently, this effectively features the spin-orbit interaction.  This unconventional type of Kondo hybridization gives rise to an effective ferromagnetic Ruderman-Kittel-Kasuya-Yosida (RKKY) interaction among the localized $f$-orbitals via perturbation theory, leading to spin-triplet resonating-valence-bond (RVB) pairing between f-electrons with $p \pm ip$-wave gap symmetry.  There exists competition and collaboration between the Kondo hybridization and the ferromagnetic RKKY coupling. Via a static mean-field approach, we explore the phase diagram of this model. Apart from a Kondo-dominating and a RVB-dominating phase, we find a Kondo triplet RVB coexisting phase in the intermediate range of the Kondo to RKKY coupling ratio. This is a superconducting phase with spin-triplet $p \pm ip$-wave pairing gap. We further show that this co-existing phase is a time-reversal invariant topological superconducting state which supports helical Majorana zero modes at the edges of a nano-strip system. Our results provide an unique example of topological superconductivity arising from Kondo correlations, and are relevant for the possible topological Kondo superconducting state in heavy-fermion compounds.
Spin-triplet $p$-wave superconductors are promising candidates for topological superconductors. They have been proposed in various heterostructures where a material with strong spin-orbit interaction is coupled to a conventional $s$-wave superconductor by proximity effect. However, topological superconductors existing in nature and   driven purely by  strong electron correlations are yet to be studied. Here we propose a  realization of such a system in a class of Kondo lattice materials in the absence of spin-orbit coupling and  proximity effect. Therein, the odd-parity Kondo hybridization mediates ferromagnetic  spin-spin coupling and leads to spin-triplet resonant-valence-bond ($t$-RVB) pairing between local moments. Spin-triplet $p\pm i p^\prime$-wave topological superconductivity is reached when Kondo effect co-exists with $t$-RVB. We identify the topological nature by the non-trivial topological invariant and the Majorana fermions at edges. Our results offer a comprehensive understanding of experimental observations on UTe$_2$, a U-based ferromagnetic heavy-electron superconductor.
%Experimental implications of our results on the $U$-based ferromagnetic heavy-electron superconductors with Kondo correlations, in particular UTe$_2$, is discussed.  
\end{abstract}

\maketitle

\section{Introduction}
%{\color{blue}(Intro. to topological TI in strongly correlated electrons)} 
Searching for topological superconductors (TSc) and the corresponding self-dual charge neutral Majorana zero modes associated with their excitations at edges has become one of the central problem in condensed matter physics \cite{Liang-RMP-2011,Alicea-MF-2012}. Theoretical proposals and experimental realizations of TSc are mostly heterostructure combining strong spin-orbit coupled materials and conventional superconductors by proximity effect \cite{Sau-PRL-MF,Oreg-PRL-MF,Paaske-PRL-MF}. The emergence of the topological edge states in such systems can be explained in terms of the single-particle band structure without considering many-body electron correlations. Recently, the search for topological phases of matter  has focused on a more intriguing class of materials that exist in nature.  Their topological properties are driven by strong electron correlations instead of the proximity effect. Kondo effect, describing the screening of a local spin moment by conduction electrons, is a well-known strong correlation between electrons existing in heavy electron compounds. The Kondo-mediated topological phases of matter have been studied in the context of topological Kondo insulators \cite{dzero-Ann-TKI,dzero-TKI-PRB,dzero-TKI-PRL} and topological Kondo semi-metals \cite{Weyl-kondo-PNAS}, where the topological properties are driven by either the odd-parity Kondo hybridization or by the Kondo hybridization with strong spin-orbit coupling.
%Searching for new quantum states with non-trivial topological properties has continued to be one of the central issues in condensed matter physics. One of the well-known examples is the time-reversal invariant topological insulator/superconductor in which the strong spin-orbit coupling/$p$-wave triplet pairing potential leads to a topologically distinct ground-state wave function in comparison with the topologically trivial vacuum, giving rise to helical gapless edge (or surface) states with gaped bulk band structure  \cite{Qi-RMP-2009,Hasan-RMP}. The emergence of these edge states and the associated topologically trivial-to-nontrivial transition in a system can be explained in terms of the band structure (single-particle) effect without considering (many-body) interactions. An even more intriguing class of topological states and the associated topological phase transition is driven by strongly electronic correlation, and exploring such ``interacting'' topological states has become a current interest \cite{Rachel-2018-RPP}. 

%{\color{blue}(Introduce TKI's}
 
Spin-triplet $p$-wave superconductors are known to be the prime candidates for TSc. However, they are scarce in nature. While it is still debatable for SrRu$_2$O$_4$ \cite{Mackenzie-RMP-SrRuO,Maeno-JPSJ-SrRuO,Kallin-JPCM-2009}, more convincing evidence for $p$-wave triplet superconductivity  was observed in noncentrosymmetric superconductor BiPd from  phase-sensitive measurement \cite{CLChen-BiPd-PRL}. More recently, signatures of triplet chiral $p$-wave  superconductivity were observed in heavy-electron Kondo lattice compound UTe$_2$ at the edge of ferromagnetism, possibly marking the first example of topological superconductor induced by the strongly correlated  Kondo effect \cite{Ran-FM-UTe,Ran-UTe-Nature-ExtremeMag,Aoki-2019-JPSJ,Jiao-2020-UTe2}.  
 
 Motivated by these discoveries, in this paper, we propose  a distinct class of triplet $p$-wave superconductors in the absence of spin-orbit coupling or proximity effect/heterostructure \cite{Kim-Kondo-Kitaev} in a two-dimensional Kondo lattice model driven by odd-parity Kondo hybridization. We start from the Anderson lattice model (ALM) with odd-parity hybridization, which occurs between $d$- and $f$-orbital electrons in various heavy-fermion compounds \cite{dzero-Ann-TKI,dzero-TKI-PRB,dzero-TKI-PRL}. Via the Schrieffer-Wolff transformation \cite{SW-transformation,hewson1997kondo}, we  derive an effective Kondo lattice model with odd-parity hybridization. Furthermore, by integrating out the conduction electron degrees of freedom, an effective ferromagnetic RKKY interaction is generated. We explore the mean-field phase diagram of this ferromagnetic Kondo-Heisenberg model.  In the fermionic mean-field approach, the ferromagnetic RKKY coupling describes the $p$-wave ($S_z=\pm 1$) $t$-RVB spin-liquid state. A time-reversal invariant topological superconducting phase is reached when the Kondo effect co-exists with the $p$-wave $t$-RVB order parameter. The topological nature of this superconducting phase is manifested by the non-trivial $Z_2$ topological  Chern number of the bulk band and by the existence of helical Majorana zero modes at the edges of a finite-sized ribbon. Our results offer a qualitative and some quantitative understanding of the spin-triplet superconductivity recently observed in UTe$_2$ (see Discussions). 
\section{Model}
\subsection{Anderson lattice model with odd-parity hybridization}
We start with the odd-parity Anderson lattice model  (ALM) on a two-dimensional (2D) square lattice, which has been shown to exhibit topologically non-trivial states \cite{dzero-Ann-TKI,dzero-TKI-PRB,dzero-TKI-PRL}:
\begin{align}
H_{PAM}= H_c +H_f+ H_{cf} , \label{eq:PAM}
\end{align}
where $H_c = \sum_{\mbd{k},\sigma = \uparrow,\downarrow} \varepsilon_\mbd{k} c^\dagger_{\mbd{k}\sigma}c_{\mbd{k}\sigma}$ describes the hopping of electrons in the $d$ orbits with orbital angular momentum $l=2$ and  dispersion $\varepsilon_\mbd{k}=-2t (\cos k_x + \cos k_y)-\mu$.   The Hamiltonian $H_f$ of the more localized electron in the $f$ orbits with orbital angular momentum $l=3$ is given by
\begin{align}
    H_f = \sum_{i,\sigma}\left[ \varepsilon_f f^\dagger_{i\sigma}f_{i\sigma}+\frac{U}{2}n_{i\sigma}^{f}n_{i,-\sigma}^{f}\right],
\end{align}
 where $\varepsilon_f$ denote the energy level of the $f$-electron, and $U$ is the repulsive on-site Coulomb potential (the Hubbard-$U$ term). Hybridization of the local and conduction electrons is described by
 \begin{align}
     H_{cf} = \sum_{\langle i,j\rangle}\sum_{ \sigma,\sigma^\prime = \uparrow\downarrow} V_{ij}^{\sigma \sigma^\prime} c^\dagger_{i\sigma}f_{j\sigma^\prime}+H.c..
 \end{align}
%In general, we may consider a site- and spin-dependent hybridization, $V_{ij}^{\sigma \sigma^\prime}$. distinct from the onsite and spin-conserving one of the well-known Anderson model. 
To conserve the parity symmetry of hybridization between electrons with their angular momentum quantum numbers differing by one, $V_{ij}^{\sigma\sigma^\prime}$ have to be odd under parity transformation. This restriction results in the hybridization having to depend on sites and spins \cite{dzero-Ann-TKI,dzero-TKI-PRB,dzero-TKI-PRL}:
 \begin{align}
 V_{ij}^{\sigma\sigma^\prime} \equiv V_{\hat{\alpha}}^{\sigma\sigma^\prime} =  iV\nu_{\hat{\alpha}}\sigma_{{\alpha}}^{\sigma\sigma^{\prime}},
 \label{eq:V-ij}
 \end{align}
 distinct from the well-known onsite and spin-conserving  Anderson hybridization.  In Eq. (\ref{eq:V-ij}), $\nu_{ij}$ satisfies $\nu_{ij} \equiv \nu_{\hat{\alpha}} = -\nu_{ji}$ with  $\hat{\alpha} \equiv i-j \in \ \hat{x}, \hat{y} \, (\alpha \in x,y)$  on a  2D square lattice, and $\sigma_{\alpha}$ denotes the Pauli matrix of the $\alpha$ component.
 
 \subsection{The effective odd-parity ferromagnetic Kondo lattice model}
 
 \begin{figure}[t]
     \centering
     \includegraphics[width = 0.45 \textwidth]{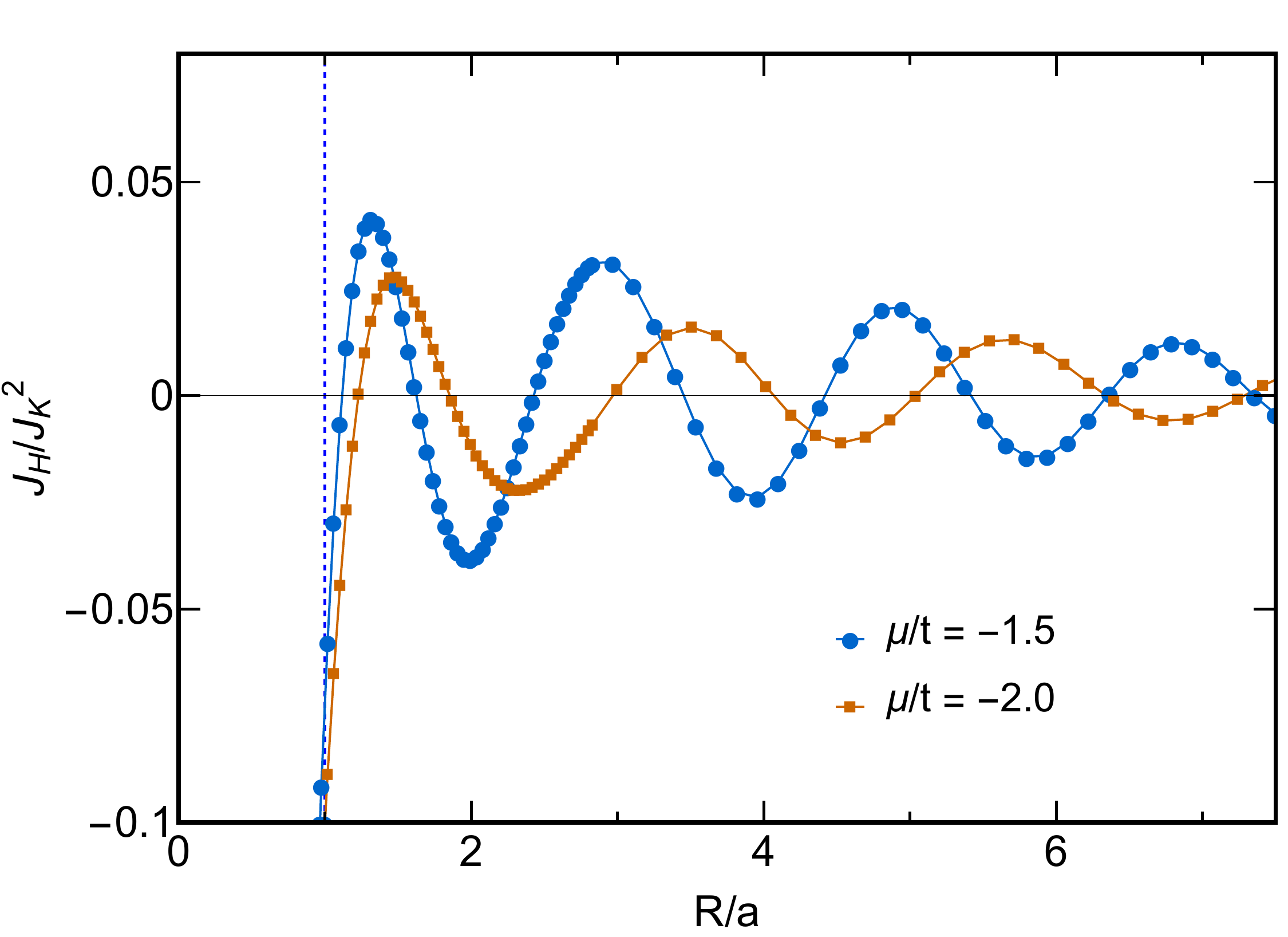}
     \caption{The effective RKKY coupling $J_H$ (normalized with $J_K^2$) as a function of $R/a$ for different chemical potentials $\mu$. $J_H$ is computed by Eq. (\ref{eq:J_H}) with $\mbd{R}_{ij} \parallel (1,1) $ and $a=1$ being chosen.}
     \label{fig:JH}
 \end{figure} 
 
\added{In this paper, we focus on the competition of the Kondo and the magnetic interaction among impurities--the Doniach scenario \cite{Doniach-scenario}. We, therefore, derive the effective Kondo-Heisenberg lattice Hamiltonian from ALM in the Kondo limit  where the vacant and doubly-occupied states are projected out from the entire Hilbert space, namely $1 = \sum_\sigma  f^\dagger_{i\sigma}f_{i\sigma}$. The low-energy effective Kondo term  from the odd-parity ALM of Eq. (\ref{eq:PAM}) can be derived by applying the Schrieffer-Wolff transformation (SWT) \cite{SW-transformation,hewson1997kondo,ETH-Phd-thesis}, yielding}

  \begin{align}
      H_K =  (-J_K) & \sum_i\sum_{\sigma\sigma^{\prime}}\sum_{\sigma^{\prime\prime}\sigma^{\prime\prime\prime}}\sum_{\alpha,\alpha^\prime}\left(i\nu_{\hat{\alpha}}\sigma_{\alpha}^{\sigma\sigma^{\prime}}c_{i+\hat{\alpha},\sigma}^{\dagger}f_{i\sigma^{\prime}}\right)
      \nn
      & \times \left(i\nu_{\hat{\alpha}^{\prime}}\sigma_{\alpha^{\prime}}^{\sigma^{\prime\prime}\sigma^{\prime\prime\prime}}f_{i\sigma^{\prime\prime}}^{\dagger}c_{i-\hat{\alpha}^{\prime},\sigma^{\prime\prime\prime}}\right)
      \label{eq:HK}
  \end{align}
  
\added{with $J_{K}=\frac{V^{2}}{U+\varepsilon_{f}-\varepsilon_{F}}+\frac{V^{2}}{\varepsilon_{F}-\varepsilon_{f}}>0$ (see Appendix \ref{app:SWT}). The Kondo-like term of Eq. (\ref{eq:HK}) describes the screening of an impurity by its neighboring conduction electrons, distinct from the conventional (on-site) Kondo term.}  \deleted{Via using the Schrieffer-Wolff transformation (SWT) to generate the Kondo term and via applying the perturbative approach to generate the antiferromagnetic RKKY interaction,}  

\replaced{Here, we go beyond the topological Kondo insulating phase by  further deriving the magnetic RKKY interaction among the local $f$-fermions.  By perturbatively expanding the Kondo term to second order  \cite{rkky-kittel,rkky-vleck,ETH-Phd-thesis}}{Applying the second-order perturbation in $J_K$}, we obtain the effective RKKY-like interaction between the local $f$ fermions $f_{i\sigma}$,  
  \begin{align}
      H_J =& \sum_{ i,j }\sum_{\sigma ,\sigma^\prime} J_{ij}f_{i\sigma}^{\dagger}f_{j\sigma^\prime}^{\dagger}f_{j\sigma}f_{i\sigma^\prime}\nn
      =& \sum_{\left\langle i,j\right\rangle }J_{ij}\left(f_{i\uparrow}^{\dagger}f_{j\uparrow}^{\dagger}f_{j\uparrow}f_{i\uparrow}+f_{i\downarrow}^{\dagger}f_{j\downarrow}^{\dagger}f_{j\downarrow}f_{i\downarrow}\right)\nn
    &+ \sum_{\left\langle i,j\right\rangle }\frac{J_{ij}}{2}\left(f_{i\uparrow}^{\dagger}f_{j\downarrow}^{\dagger}+f_{i\downarrow}^{\dagger}f_{j\uparrow}^{\dagger}\right)\left(f_{j\downarrow}f_{i\uparrow}+f_{j\uparrow}f_{i\downarrow}\right) \nn 
    & - \sum_{\left\langle i,j\right\rangle }\frac{J_{ij}}{2} \left(f_{i\uparrow}^{\dagger}f_{j\downarrow}^{\dagger}-f_{i\downarrow}^{\dagger}f_{j\uparrow}^{\dagger}\right)\left(f_{j\downarrow}f_{i\uparrow}-f_{j\uparrow}f_{i\downarrow}\right),
      \label{eq:HJ}
  \end{align}
 where 
 \begin{align}
     J_{ij} \equiv J_H(& R)=\frac{16J_{K}^{2}}{\mathcal{N}_{s}^{2}}\sum_{\varepsilon_{\mbd{k}}<\mu}\sum_{\varepsilon_{\mbd{k}^{\prime\prime}}>\mu} \frac{e^{i\left(\mbd{k}-\mbd{k}^{\prime\prime}\right)\cdot\mbd{R}_{ij}}}{\varepsilon_{\mbd{k}}-\varepsilon_{\mbd{k}^{\prime\prime}}} \nn
     &\times \left(\sin^{2}k_{x}+\sin^{2}k_{y}\right)\left(\sin^{2}k_{x}^{\prime\prime}+\sin^{2}k_{y}^{\prime\prime}\right)
     \label{eq:J_H}
 \end{align}
denotes the effective coupling of the spinons of sites $i$ and $j$ with $R \equiv |\mbd{R}_{ij}| \equiv |\mbd{r}_i -\mbd{r}_j|$. \added{The $H_J$ term of Eq. (\ref{eq:HJ}) can be  re-expressed as a linear combination of a spinon pair wave function with total spin $S  = 0$ (spin-singlet) and $S = 1$ (spin-triplet). Note that  the associated effective spinon coupling of the spin-triplet channel is opposite to that of the spin-singlet}. \added{When $H_J$ is expressed in terms of fermion pair with different spins,   Eq. (\ref{eq:HJ}) is reminiscent of the conventional Heisenberg interaction $\mbd{S}_{i}\cdot \mbd{S}_{j} = -\frac{1}{2}\left(f_{i\uparrow}^{\dagger}f_{j\downarrow}^{\dagger}-f_{i\downarrow}^{\dagger}f_{j\uparrow}^{\dagger}\right)\left(f_{i\downarrow}f_{j\uparrow}-f_{i\uparrow}f_{j\downarrow}\right)+\frac{1}{4}n_{i}^{f}n_{j}^{f}$, except for the difference in the constant coefficients of the pair operators.} \deleted{As compared to the conventional RKKY coupling obtained from the perturbation of the Kondo term with constant Anderson hybridization, we expect that} \added{As expected, the RKKY coupling $J_{ij}$ in Eq. (\ref{eq:J_H}) shows an oscillatory behavior in  $R$, accompanied by a decrease in its magnitude with increasing  $R$}\deleted{of the coupling strength $J_{ij}$ in Eq. (\ref{eq:J_H}) tends to}, \added{similar to the behavior of the conventional RKKY coupling}. \deleted{Here,} \added{Due to the rapid attenuation of  $J_{ij}$, we only consider} \deleted{We thus only retain} \added{the dominated nearest-neighbor interaction and  assume $J_{ij}$ to be  spatially  homogeneous, i.e.} \deleted{with} $J_{ij} \to J (R = a) \equiv J_H$\deleted{denoted the coupling of the nearest-neighbor sites}. \added{Furthermore, when $R = a$, we find the effective RKKY coupling is attractive (or of the ferromagnetic type), i.e., $J_H<0$ (see Fig. \ref{fig:JH}), which energetically favors the spin-triplet pairing of spinons. On the other hand, the effective RKKY coupling in the spin-singlet channel shows repulsive interaction and  can be neglected here since it is not energetically favorable.} Lastly, on a two-dimensional lattice, the \added{triplet spin state  $|\uparrow\downarrow\rangle + | \downarrow \uparrow \rangle$ does not exist since the corresponding structure factor is proportional to $k_z$, and $k_z = 0$ is fixed here}. Therefore, based on the above arguments, only the equal-spin states,  $|\uparrow \uparrow \rangle$ and $|\downarrow \downarrow \rangle$, survive, and the $H_J$ term is reduced to
 \begin{align}
     H_{J}\approx-\left|J_{H}\right|\sum_{\left\langle i,j \right\rangle }\left(f_{i\uparrow}^{\dagger}f_{j\uparrow}^{\dagger}f_{j\uparrow}f_{i\uparrow}+f_{i\downarrow}^{\dagger}f_{j\downarrow}^{\dagger}f_{j\downarrow}f_{i\downarrow}\right).
     \label{eq:HJ-triplet}
 \end{align}

Combining $H_K$ and $H_J$ of Eqs. (\ref{eq:HK}), (\ref{eq:HJ}) and (\ref{eq:HJ-triplet}), the  effective Kondo-Heisenberg lattice model with odd-parity Kondo hybridization reads $H_{FKH}=H_{0}+H_\lambda+H_{K}+H_{J}$. Here, $H_\lambda =- \sum_{i} i\lambda_i \left[ \sum_\sigma (f^\dagger_{i\sigma}f_{i\sigma})-1 \right]$ enforces the singly occupied local $f$-spinons with $\lambda_i$ being the Lagrange multiplier. \deleted{$H_K$ and $H_J$ describes the Kondo screening and the antiferromagnetic exchange coupling of the impurity spins.} \added{The Hamiltonian $H_{FKH}$ offers a platform for discovering a distinct class of topological superconducting states induced by electron correlations  via collaboration between the ferromagnetic RKKY coupling and the Kondo effect. To facilitate our numerical calculations of the mean-field phase diagram, we treat $J_K$ and $J_H$ as independent couplings here since it is more convenient to explore the phase diagram by tuning the ratio of $J_K/J_H$ \cite{RMP-Kirchner,sm-phase-PNAS}. In experiments, varying the non-thermal parameter can be expected to follow a certain trajectory of $J_K/J_H$ in the phase diagram.}

 %%%%%%%%%%%%%%%%%%%%%%%%%%%%%%%%%%%
 %%%%%%%%%%%%%%%%%%%%%%%%%%%%%%%%%%%
\section{Mean-field treatment of the effective Kondo-Heisenberg-like model}

\begin{figure}[t]
    \centering
    \includegraphics[width=0.45\textwidth]{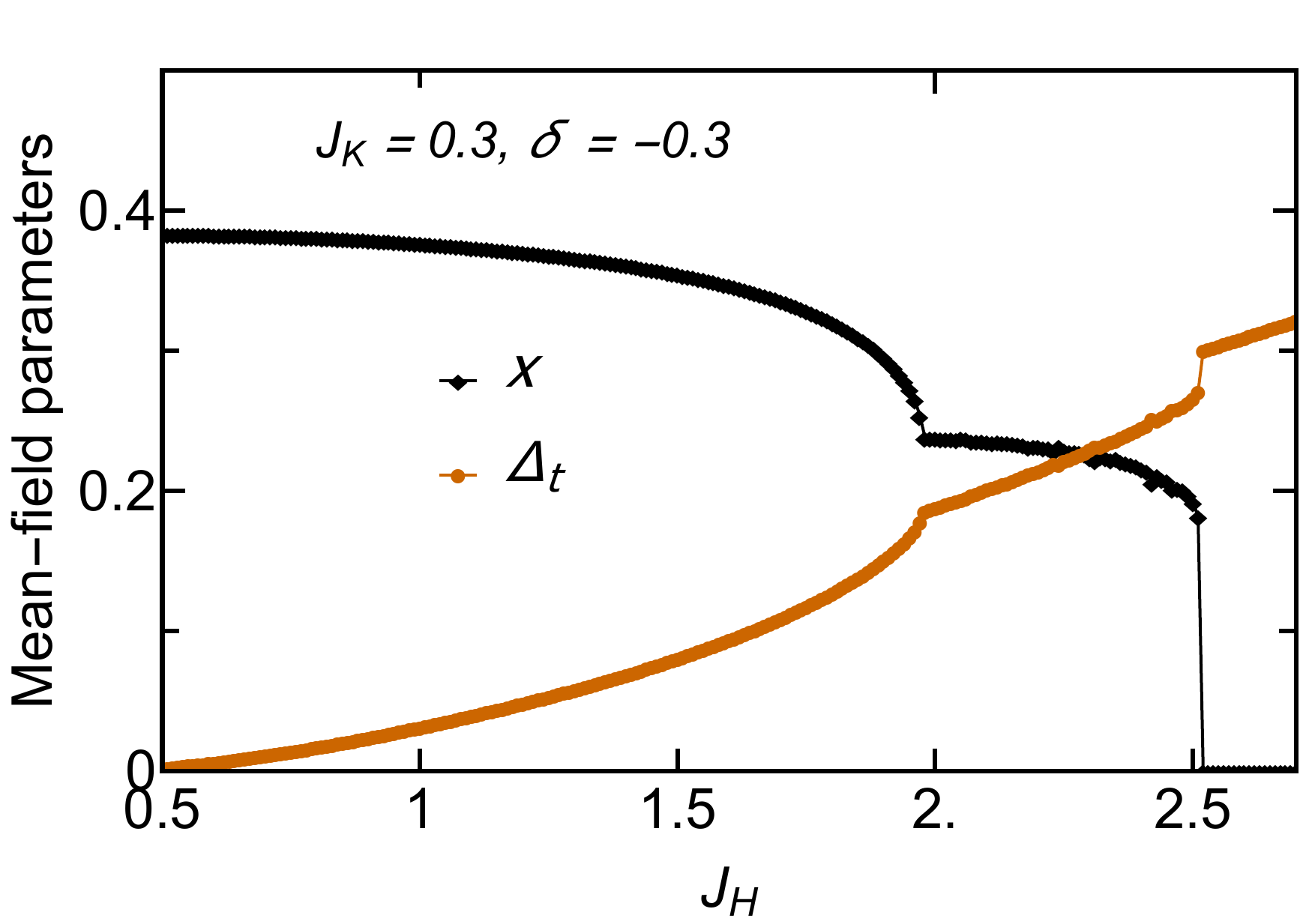}
    \caption{The zero-temperature mean-field solutions of $t$-RVB order parameter $\Delta_t$ (brown) and the Kondo correlation $x$ (black) as a function of $J_H$. We fix $J_K = 0.3$ and doping of the conduction band $\delta = -0.3$ ($30$ percent hole doping). Without loss of generality, \added{we set} $t = 1$. This plot reveals a (co-existing) superconducting ground state with  $x \neq 0,\,\Delta_t \neq 0$ for $0 < J_H \lesssim 2.5$ and a pure $t$-RVB phase where  $x = 0,\,\Delta_t \neq 0$ when $J_H \gtrsim 2.52$. A pure Kondo phase ($x \neq 0,\,\Delta_t = 0$) only exists at $J_H = 0$.}
    \label{GS}
\end{figure}

\added{We now employ a mean-field analysis on the above effective Kondo-Heisenberg-like Hamiltonian with an effective ferromagnetic RKKY interaction and odd-parity Kondo hybridization.} 

Via performing Hubbard-Stratonovich transformation, $H_K$ and $H_J$ of Eqs. (\ref{eq:HK}) and (\ref{eq:HJ}) can be factorized as
\begin{align}
    &H_K\to  \sum_{i,\alpha}\sum_{\sigma\sigma^{\prime}} \left[\chi^\dagger_i
       \left(i\nu_{\hat{\alpha}}\sigma_{\alpha}^{\sigma\sigma^{\prime}}f_{i\sigma}^{\dagger}c_{i-\hat{\alpha},\sigma^{\prime}}\right) +H.c. \right]\nn
       & \qquad\qquad+ \sum_i \frac{|\chi_i|^2}{J_K}, \nn
& H_J \to \sum_{\left\langle i,j\right\rangle }\left[\Delta^{\uparrow}_t(i,j)f_{i\uparrow}^{\dagger}f_{j\uparrow}^{\dagger}+\Delta^{\downarrow}_t(i,j)f_{i\downarrow}^{\dagger}f_{j\downarrow}^{\dagger}+H.c.\right] \nn
& \qquad\qquad+\sum_{\langle i,j\rangle}\frac{\left|\Delta^{\uparrow}_t (i,j) \right|^{2}+\left|\Delta^\downarrow_t (i,j) \right|^{2}}{J_{H}}
\label{eq: HS}
\end{align}

where the mean-field values of the bosonic Hubbard-Stratonovich fields, $\chi_i$ and $\Delta^{\sigma}_t (i,j)$ ($\sigma = \uparrow,\downarrow$), represent the order parameters of the Kondo correlation and the $S_z = \pm 1$ spin-triplet RVB bonds between two adjacent up/down spins, respectively. \deleted{In $H_J$, we only retain the components with spin-triplet pairings. We argue that the spin-singlet channel can be omitted if the effective Heisenberg coupling $J_H$ is negative, which makes the spin-singlet pairing to be unstable (see  \sia).}

To describe the Kondo-screened  Fermi-liquid state, we allow the $\chi_i$ field to acquire uniformly Bose condensation over the real space; hence, $\chi_i$ can be expressed as $\chi_i\to x + \hat{\chi}_i$ with $x= (-J_K/\mathcal{N}_s)\sum_{i\sigma\sigma^\prime\alpha}\langle i\nu_{\hat{\alpha}}\sigma_{\alpha}^{\sigma\sigma^{\prime}}f_{i\sigma}^{\dagger}c_{i-\hat{\alpha},\sigma^{\prime}} \rangle$ being the Bose-condensed stiffness of $\chi_i$ while $\hat{\chi}_i$ represents its fluctuations. The mean-field order parameter of the $t$RVB is given by $\Delta_t^{\sigma}= (-J_H/4\mathcal{N}_s)\sum_{\langle i,j\rangle}\langle f_{j\sigma}f_{i\sigma}\rangle$. \deleted{To acquire the mean-field model of the Hamiltonian,} Since the ferromagnetic coupling is expected to favor spin-triplet $p$-wave pairing similar to superfluid helium-3 \cite{Mineev-SC}, we restrict ourselves to the $p$-wave pairing, i.e., $\Delta_t^\sigma (i,j)$ here is taken the $p$-wave form, see Eqs. (\ref{eq:Delta-up}) and (\ref{eq:Delta-down}) below.  We further fix the Lagrange multiplier at the mean-field level via $i\lambda_i\to \lambda$ and neglect the fluctuations of $\lambda_i$, $\chi_i$, and $\Delta_{t}^{\sigma}$, leading to the following mean-field Kondo-Heisenberg-like Hamiltonian:

 \begin{align}
 H_{MF}&  = \sum_{\mbd{k},\sigma}\varepsilon_{\mbd{k}}c_{\mbd{k}\sigma}^{\dagger}c_{\mbd{k}\sigma} + \sum_{\mbd{k}\sigma}\lambda f_{\mbd{k}\sigma}^{\dagger}f_{\mbd{k}\sigma}\nn
 &  + \sum_{\mbd{k}}\left[V_{1\mbd{k}}f_{\mbd{k}\uparrow}^{\ast}c_{\mbd{k}\downarrow}+V_{2\mbd{k}}f_{\mbd{k}\downarrow}^{\ast}c_{\mbd{k}\uparrow}+H.c.\right]
 \nn
 & + \sum_{\mbd{k}}\left[\Delta_{\mbd{k}}^{\uparrow}f_{\mbd{k}\uparrow}^{\dagger}f_{-\mbd{k}\uparrow}^{\dagger}+\Delta_{\mbd{k}}^{\downarrow}f_{\mbd{k}\downarrow}^{\dagger}f_{-\mbd{k}\downarrow}^{\dagger}+H.c.\right]\nn
 & +\frac{8\mathcal{N}_{s}\Delta_{t}^{2}}{J_{H}}+\frac{\mathcal{N}_{s}x^{2}}{J_{K}}-\mathcal{N}_{s}\lambda,
 \label{eq:MF-H}
 \end{align}
 
where $V_{1\mbd{k}}=2x\left(\sin k_{x}-i\sin k_{y}\right)$ and $ V_{2\mbd{k}}=2x\left(\sin k_{x}+i\sin k_{y}\right)$. The Fourier transformation for the second-quantized operator is defined as $\psi_{i\sigma}=\frac{1}{\sqrt{\mathcal{N}_s}}\sum_{\mbd{k}}e^{-i\mbd{k}\cdot \mbd{r}_i}\psi_{\mbd{k}\sigma}$. Note that the mean-field Kondo term of Eq. (\ref{eq:MF-H}) is reminiscent of the  topological Kondo insulator shown in Ref. \cite{coleman-book}.  In Eq. (\ref{eq:MF-H}), $\Delta_t^{\sigma}(\mbd{k})$  represents the gap structure of the spin-triplet $p$-wave RVB pairing in the momentum space for the spin-$\sigma$ sector, defined as $\Delta_{\mbd{k}}^{\uparrow}=\Delta_{t}\left(-\sin k_{y}-i\sin k_{x}\right)$ and $\quad \Delta_{\mbd{k}}^{\downarrow}=\Delta_{t}\left(\sin k_{y}-i\sin k_{x}\right)$ with $\Delta_t$ being denoted the mean-field pairing potential (see Appendix,  Section II). This momentum-dependent gap structure for the up- and down-spin sectors correspond to the following real-space patterns of $\Delta^\uparrow_t (i,j)$ and $\Delta^\downarrow_t(i,j)$ of Eq. (\ref{eq: HS}):
  \begin{align}
     &\Delta_t^{\uparrow}(i,j)\to  \Delta^{\uparrow}_t(i,i+\hat{x})=-\Delta^{\uparrow}_t(i,i-\hat{x})=-\Delta_{t},\nn
     &\qquad \qquad \quad \Delta^{\uparrow}_t(i,i+\hat{y})=-\Delta^{\uparrow}_t(i,i-\hat{y})=i\Delta_{t},
     \label{eq:Delta-up}
 \end{align}
 and 
 \begin{align}
     \Delta_t^{\downarrow}(i,j)\to& \Delta^{\downarrow}_t(i,i+\hat{x})=-\Delta^{\downarrow}_t(i,i-\hat{x})=-\Delta_{t},\nn & \Delta^{\downarrow}_t(i,i+\hat{y})=-\Delta^{\downarrow}_t(i,i-\hat{y})=-i\Delta_{t}.
     \label{eq:Delta-down}
 \end{align}
 
Choosing $\Psi_{\mbd{k}} = \left( \phi_{A\mbd{k}}, \phi_{B\mbd{k}} \right)^T$ with the Nambu spinors defined by  $\phi_{A\mbd{k}}=\left(c_{\mbd{k}\uparrow},\,\,c_{-\mbd{k}\uparrow}^{\dagger},\,\,f_{\mbd{k}\downarrow},\,\,f_{-\mbd{k}\downarrow}^{\dagger}\right)^T$ and  $\phi_{B\mbd{k}}=\left(c_{\mbd{k}\downarrow},\,\,c_{-\mbd{k}\downarrow}^{\dagger},\,\,f_{\mbd{k}\uparrow},\,\,f_{-\mbd{k}\uparrow}^{\dagger}\right)^T$,  the mean-field Hamiltonian $H_{MF} = \sum_{\mbd{k}}\Psi_{\mbd{k}}^{\dagger}\mathcal{H}_{\mbd{k}}\Psi_{\mbd{k}} + \mathcal{C}$ can be expressed as a summation of two decoupled $4 \times 4$ matrices as follows
 \begin{align}
     H_{MF}& = H_A + H_B  + \mathcal{C}, \nn
     H_{A(B)}& =  \sum_{\mbd{k}}\phi_{A(B)\mbd{k}}^{\dagger}\mathcal{H}_{\mbd{k}}^{A(B)}\phi_{A(B)\mbd{k}}
     \label{eq:H-MF}
 \end{align}
 
 with $\mathcal{C} \equiv \sum_\mbd{k} \varepsilon_\mbd{k} + \frac{8\mathcal{N}_{s}\Delta_{t}^{2}}{J_{H}}+\frac{\mathcal{N}_{s}x^{2}}{J_{K}}$, and
 
 \begin{align}
     \mathcal{H}_{\mbd{k}}^{A} & =\begin{pmatrix}
\frac{\varepsilon_{\mbd{k}}}{2} 	& 0	& \frac{V_{2\mbd{k}}^{\ast}}{2}	& 0				\\
0 & -\frac{\varepsilon_{\mbd{k}}}{2} 	& 0	& \frac{V_{2\mbd{k}}}{2}				\\
\frac{V_{2\mbd{k}}}{2} 		& 0	& \frac{\lambda}{2} 		& \Delta_{\mbd{k}}^{\downarrow}	\\
0 & \frac{V_{2\mbd{k}}^{\ast}}{2}	& \Delta_{\mbd{k}}^{\downarrow\ast} 	& -\frac{\lambda}{2}
\end{pmatrix},\,\,\\
\mathcal{H}_{\mbd{k}}^{B} & =\begin{pmatrix}
\frac{\varepsilon_{\mbd{k}}}{2} 	& 0	& \frac{V_{1\mbd{k}}^{\ast}}{2}	& 0				\\
0 & -\frac{\varepsilon_{\mbd{k}}}{2} 	& 0	& \frac{V_{1\mbd{k}}}{2}				\\
\frac{V_{1\mbd{k}}}{2} 		& 0	& \frac{\lambda}{2} 		& \Delta_{\mbd{k}}^{\uparrow}	\\
0 & \frac{V_{1\mbd{k}}^{\ast}}{2}	& \Delta_{\mbd{k}}^{\uparrow\ast} 	& -\frac{\lambda}{2}.
\end{pmatrix}
 \end{align}
 
The Hamiltonian Eq. (\ref{eq:H-MF})  possesses time-reversal symmetry: $\mathcal{H}_{A}$ and $\mathcal{H}_{B}$ constitute the time-reversal partner of each other, i.e. $\Theta \mathcal{H}_{A (B)} \Theta^{-1} = \mathcal{H}_{B(A)}$ where the time-reversal operator $\Theta = \rho^0 \otimes (-i\sigma^{y})K$ with $\sigma^y $ being the $y$-component Pauli matrix on the spin subspace, $\rho^0$ being a $2\times 2 $ identity matrix on the orbital subspace while $K$ being the complex-conjugate operator. Under time-reversal transformation, the spin and quasi-momentum of conduction ($c$) and pseudofermion ($f$) operators are flipped: $ \left(c_{\bm{k}\uparrow},\, c_{\bm{k}\downarrow}, \, f_{\bm{k}\uparrow},\, f_{\bm{k}\downarrow}\right) \stackrel{\Theta}{\longrightarrow}  \left(c_{-\bm{k}\downarrow},\, -c_{-\bm{k}\uparrow}, \, f_{-\bm{k}\downarrow},\, -f_{-\bm{k}\uparrow}\right)$. Meanwhile, our Hamiltonian respects charge-conjugation (particle-hole) symmetry:  $\mathcal{P}\mathcal{H}_{\bm{k}}\mathcal{P}^{-1} = -\mathcal{H}_{-\bm{k}}$ where  $\mathcal{P} \equiv \tau^x K$ is the particle-hole operator  with $\tau_x$ being the $x$-component of the Pauli matrices on the particle-hole basis. Due to the odd-parity $p\pm ip^\prime$ RVB pairing of our model, the parity symmetry is broken here. Thus, our model Eq. (\ref{eq:H-MF}) belongs to the DIII class of topological symmetry \cite{Shinsei-PRB-classification}.  \deleted{Therefore, instead of using the ($8 \times 8$) $\mathcal{H}_{\mbd{k}}$ matrix for self-consistent calculations, we can consider its ($4 \times 4$) components, $\mathcal{H}_{\mbd{k}}^{A}$ and $\mathcal{H}_{\mbd{k}}^{B}$, separately.}

\section{Results}
\subsection{Mean-field phase diagram}
\label{sec:MFPD}

The \added{mean-field} ground states \replaced{are}{is} determined by minimizing the mean-field \replaced{free energy per site $\mathcal{F}_{MF} = \frac{\mathcal{C}}{\mathcal{N}_s} - \frac{ k_B T}{\mathcal{N}_s} \sum_{n\bm{k}}\ln \left[ 1+\exp \left( - \frac{E_{n\bm{k}}}{k_B T}\right) \right] $}{ground state energy of $H_{MF}$} \added{with respect to the mean-field variables $q = (\lambda, \, x,\, \Delta_t)$, i.e. $\partial \mathcal{F}_{MF}/\partial q_i = 0$.} \added{ Here, $E_{n\mbd{k}} < 0$ is the $n$-th band of $\mathcal{H}_{\mbd{k}}$. The chemical potential $\mu$ is determined by the relation $\partial \mathcal{F}_{MF}/\partial \mu = -(1+\delta)$ with $\delta$ being the chemical doping of the $c$-electrons for which $\delta  \lesseqqgtr 0 $ is for $p/\text{un}/n-$ doped (half-filling corresponds to $\delta = 0$).} This leads to the following saddle-point equations at zero temperature, 

 \begin{align}
   & \frac{1}{\mathcal{N}_s}\sum_{n\bm{k}}\frac{\partial E_{n\bm{k}}}{\partial x} + \frac{2x}{J_K} = 0, \nn
   & \frac{1}{\mathcal{N}_s}\sum_{n\bm{k}}\frac{\partial E_{n\bm{k}}}{\partial \Delta_t} + \frac{16 \Delta_t}{J_H} = 0, \nn
   & \frac{1}{\mathcal{N}_s}\sum_{n\bm{k}}\frac{\partial E_{n\bm{k}}}{\partial \lambda}  = 0, \nn
   &  \frac{1}{\mathcal{N}_s}\sum_{n\bm{k}}\frac{\partial E_{n\bm{k}}}{\partial \mu} +\delta = 0 .
   \label{eq:self-const-eq}
 \end{align}
 
 %\begin{align}
  %     1 & = \frac{1}{\mathcal{N}_s}\sum_{\mbd{k}\sigma}\langle f^\dagger_{\mbd{k}\sigma}f_{\mbd{k}\sigma} \rangle ,\nn
       %x&=\text{Re}\left[-\frac{2J_{K}}{\mathcal{N}_{s}}\sum_{\mbd{k},\alpha}\sum_{\sigma,\sigma^{\prime}}\sigma_{\alpha}^{\sigma\sigma^{\prime}}\sin k_{\alpha}\left\langle f_{\mbd{k}\sigma}^{\dagger}c_{\mbd{k}\sigma^{\prime}}\right\rangle \right], \nn
     %\Delta_{t}&=-\frac{\left|J_{H}\right|}{2\mathcal{N}_{s}}\text{Re}\sum_{\mbd{k}^{\prime}}\left[\left(-\sin k_{y}^{\prime}+i\sin k_{x}^{\prime}\right)\left\langle f_{-\mbd{k}^{\prime}\uparrow}f_{\mbd{k}^{\prime}\uparrow}\right\rangle \right.  \nn
     %& \left. \quad +\left(\sin k_{y}^{\prime}+i\sin k_{x}^{\prime}\right)\left\langle f_{-\mbd{k}^{\prime}\downarrow}f_{\mbd{k}^{\prime}\downarrow}\right\rangle \right],
% \end{align}
%  Here, in 2D, the triplet superconducting state only contains the $\left|\uparrow\uparrow\right\rangle$  and $\left|\downarrow\downarrow\right\rangle$  spin states, no $\left|\uparrow\downarrow\right\rangle +\left|\downarrow\uparrow\right\rangle$ state involves as this state is generated by the $z$ component of $\mathcal{\mbd{\mathcal{L}}}_{0}(\mbd{k})$ [Eq. (3.3) in Ref. \onlinecite{Mineev-SC}]. 
\deleted{and the chemical potential for the conduction electrons can be solved via $\partial\mathcal{F}/\partial\mu_c = - n_c$,  where $\mathcal{F} = \frac{\mathcal{C}_{\mbd{k}}}{\mathcal{N}_s} - \frac{2k_BT}{\mathcal{N}_s} \sum_{n\mbd{k}}\ln\cosh\frac{E_{n\mbd{k}}}{2k_BT}$  is the free energy per site with $E_{n\mbd{k}} < 0$ being the eigenvalues of $\mathcal{H}_{\mbd{k}}$ \replaced{where}{and} $n_c = \frac{1}{\mathcal{N}_s} \sum_{i\sigma} c_{i\sigma}^\dagger c_{i\sigma}$ is the average occupation number of conduction electrons per site (i.e. $n_c > 1$ for n-doped, $= 1$ for undoped or half-filled, and $< 1$ for p-doped).}

%\section{Symmetry analysis of the mean-field model}
 %%%%%%%%%%%%%%%%%%%%%%%%%
 %%%%%%%%%%%%%%%%%%%%%%%%%
 \begin{figure}[t]
     \centering
     \includegraphics[width=0.45\textwidth]{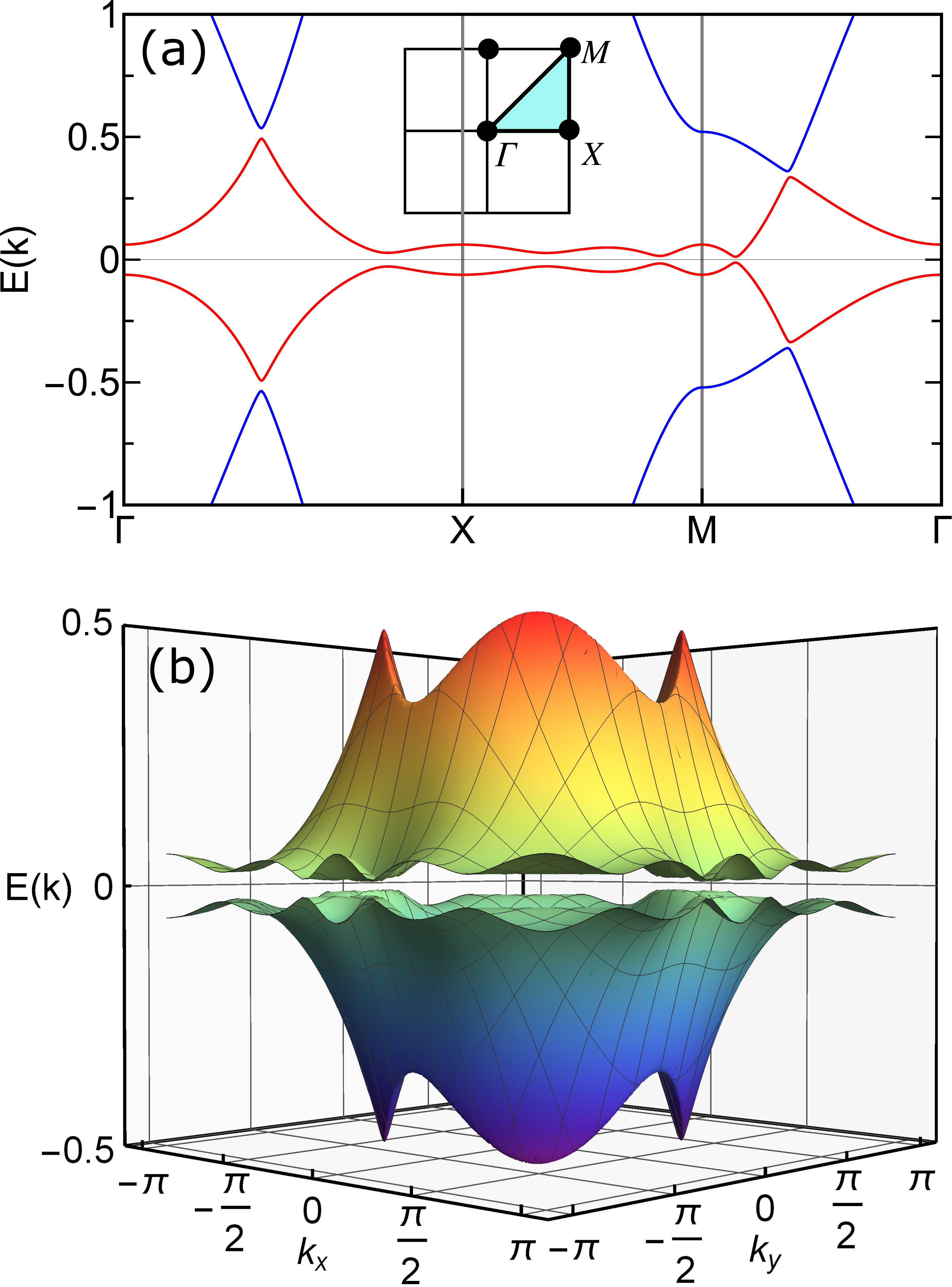}
     \caption{Figures (a) (red curves) and (b) show the bulk energy spectrum of the co-existing superconducting state near the Fermi level $\mu$. The Fermi level locates at $E(\bm k) = 0$. The coupling constants are $J_K = 0.3$ and $J_H = 1.0$. Inset of (a) displays the First Brillouin zone of a square lattice with indications of high-symmetry points $\Gamma,\, X, \,M$.}
     \label{fig:3d-disp}
 \end{figure}
 
\added{The ground-state  phase diagram (Fig. \ref{GS}) of our  model is obtained  by solving the saddle-point equations  self-consistently.} \deleted{By solving the above saddle-point equations self-consistently for both the bulk and the ribbon with $N_y = 44$ chains, we obtain the ground-state phase diagrams of the model as shown in Fig.~\ref{GS}.} \deleted{Obviously, each} \added{The} phase diagram \replaced{contains}{has} three \replaced{distinct}{separate} \added{mean-field} phases\deleted{, i.e.}: a pure Kondo phase \added{is found} at $J_H = 0$ \added{where $x \neq 0$, $\Delta_t = 0$. At the opposite limit where the RKKY interaction dominates, the ground state shows  short-range magnetic correlation with $p$-wave spin-triplet RVB   pairing ($\Delta_t \neq 0,\, x =0$). In the intermediate range of $0<J_H/J_K<(J_H/J_K)_c$, we find a Kondo-$t$RVB  co-existing (superconducting) phase with $x \neq 0$ and $\Delta_{t} \neq 0$, which can be explained via the mechanism of Kondo-stabilized spin liquid  \cite{Coleman-Andrei, sm-phase-PNAS}.  The development of superconductivity in this co-existing phase requires higher-order processes involving both the Kondo and $t$-RVB terms: the mean-field $t$-RVB pairings of the local $f$ fermions provide preformed Cooper pairs. When the Kondo hybridization field $\chi$ gets Bose-condensed ($x\neq 0$), the local fermions delocalize into the conduction band and make the preformed $t$-RVB Cooper pairs   superconduct \cite{Nevidomskyy}. These  processes can be   described by the effective mean-field Hamiltonian $H_{sc} = \sum_{\bm{k}}\left(\bar{\Delta}_{\bm{k}}^{\downarrow *}c_{-\bm{k}\downarrow
}c_{\bm{k}\downarrow} + \bar{\Delta}_{\bm{k}}^{\uparrow *}c_{-\bm{k}\uparrow
}c_{\bm{k}\uparrow} +H.c.\right)$, where the effective gap functions  take the form $\bar{\Delta}_{\bm{k}}^{\downarrow *} =V_{1\bm{k}} V_{1,-\bm{k}}\Delta_{\bm{k}}^{\uparrow*} \sim x^2 \Delta_t (\sin^2 k_x +\sin^2 k_y)(\sin k_x -i\sin k_y) $ and $\bar{\Delta}_{\bm{k}}^{\uparrow *} = V_{2\bm{k}} V_{2,-\bm{k}}\Delta_{\bm{k}}^{\downarrow*} \sim x^2 \Delta_t (\sin^2 k_x +\sin^2 k_y)(\sin k_x + i\sin k_y)$ with the size of the superconducting gap being proportional to $x^2 \Delta_t$. The superconducting gap function  $\bar{\Delta}_{\bm{k}}^{\uparrow} $ we obtained here shows a $f$-wave-like pairing symmetry on a generic anisotropic (non-circular) 2D Fermi surface.  Nevertheless, as we are taking the continuous limit of the conduction band here,  $\bar{\Delta}_{\bm{k}}^{\uparrow} $ can be expressed as a product of $s$ and $p\pm i p^\prime$ pairing orders, i.e., $\bar{\Delta}_{\bm{k}}^{\uparrow/\downarrow *} \sim k^2 (k_x \pm i k_y)$ with $k^2 \equiv k_x^2 + k_y^2$ on a circular Fermi surface but only the $p\pm i p^\prime$ component plays a role here. \added{Note that we find the co-existing superconducting state persists for an arbitrary small value of $J_H/J_K \to 0^+$. This is likely due to the overestimation of the co-existing phase at the mean-field level. Upon including fluctuations of the Kondo and  $t$-RVB order parameters beyond the mean-field level, we expect a narrower co-existing superconducting phase.}  A first-order transition similar to the results found in Refs. \cite{sm-phase-PNAS,Senthil-PRL-2003-fractionalized} is observed at the transition of the  $t$-RVB and the co-existing superconducting phases (see Fig. \ref{GS})}. \added{The bulk band structure in the co-existing superconducting state is shown in Fig. \ref{fig:3d-disp}.}
%%%%%%%%%%%%%%%%%%%%%%%%%%%%%%%%%%%%%
%%%%%%%%%%%%%%%%%%%%%%%%%%%%%%%%%%%%%
\subsection{Topological invariance}

\begin{figure}[t]
    \centering
    \includegraphics[width=0.45\textwidth]{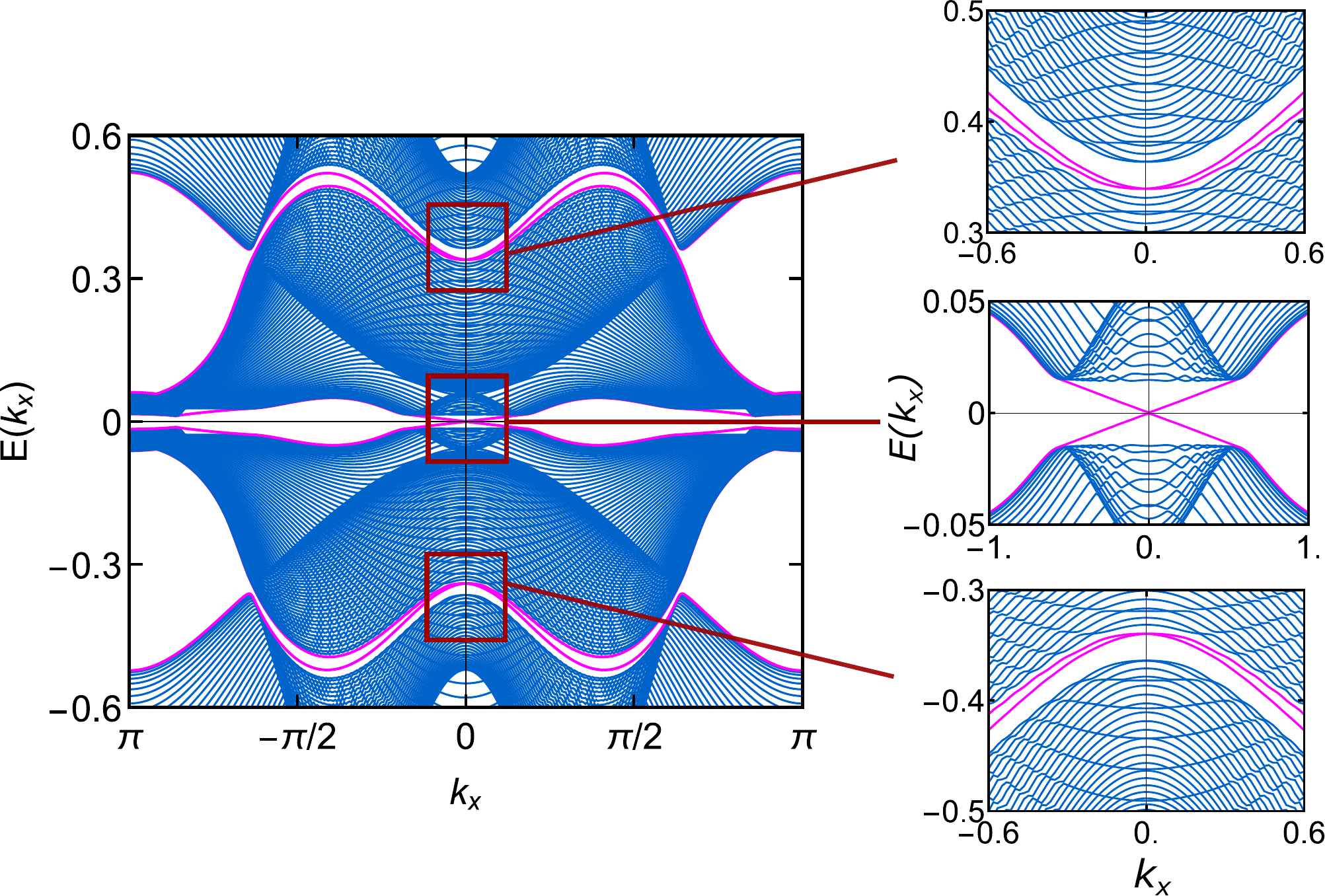}
    \caption{The left figure displays the electronic band structure of the coexisting superconductor state for a strip with $N_y = 81$ described by $\mathcal{H}_A$ at $J_K/t = 0.3$ and $J_H/t = 1.0$. Three pairs of edge states with Dirac spectra are observed near $k_x = 0$ (the pink curves). The edge states at zero energy correspond to the Majorana zero modes. Due to the time-reversal symmetry of the model, the band structure for a strip for $\mathcal{H}_B$ is identical to that of $\mathcal{H}_A$. The close-up band structures near three pairs of edge states (pink curves) on the top, middle and bottom bounded by the red squares are shown on the right figures.  }
    \label{JH_3.0_EBS_AB}
\end{figure}

We now address the topological properties of the coexisting superconducting state.  Since this system is invariant under time-reversal transformation, the bulk topological properties of the coexisting Kondo-RVB superconducting state with $p\pm i p^\prime$ spin-triplet RVB pairing can be thus characterized by  the $Z_2$ Chern number $c_T$ (or time-reversal polarization) \cite{Fu-kane-PRB-2005,Fu-kane-PRL-2007,DNSheng-PRL-2006}, given by
\begin{align}
    c_{T}=\frac{c_{A}-c_{B}}{2}
\end{align}
with $c_{I}$ $(I \in A,B)$ being the Thouless-Kohmoto-Nightingal-den Nijs (TKNN) number \cite{TKNN-PRL} of $H_I$, defined as

\begin{align}
    c_I = \frac{1}{2\pi}\int_{\bm{k}\in\text{FBZ}} d\bm{S}_{\bm{k}}\cdot  \left( \bm{\nabla}_{\bm{k}} \times \bm{\mathcal{A}}^I_{\bm{k}}\right).
\end{align}

The Berry's connection  $\bm{\mathcal{A}}^I_{\bm{k}}$ for $H_I$ is given by $ \bm{\mathcal{A}}^I_{\bm{k}} \equiv  i\sum_{n\in I} \langle u_{n\bm{k}}^{I}|\bm{\nabla_{\bm{k}}}|u_{n\bm{k}}^{I}\rangle$ with $|u_{n\bm{k}}^{I}\rangle$ being the  normalized Bloch state of the $n$-th filled band for $H^I_{\bm{k}}$. We numerically calculate the TKNN numbers \cite{Hatsugui-chern-numb}, $c_A$ and $c_B$, and find that $c_A = -c_B = 1$ in the co-existing phase, indicating a topologically non-trivial $Z_2$ Chern number $c_T = 1$. By the bulk-edge correspondence, we expect this co-existing superconducting state to support a pair of counter-propagating Majorana zero modes at the edges of a finite-sized strip. Further band structure calculations of our model on a strip in the following subsection  confirm our expectation.
%%%%%%%%%%%%%%%%%%%%%%%%%%%%%%%%%%%%
%%%%%%%%%%%%%%%%%%%%%%%%%%%%%%%%%%%%
\subsection{Edge states of the coexisting Kondo-RVB spin-triplet $p\pm ip^\prime$-wave superconducting state}

\begin{figure*}[ht]
    \centering
    \includegraphics[width=0.9\textwidth]{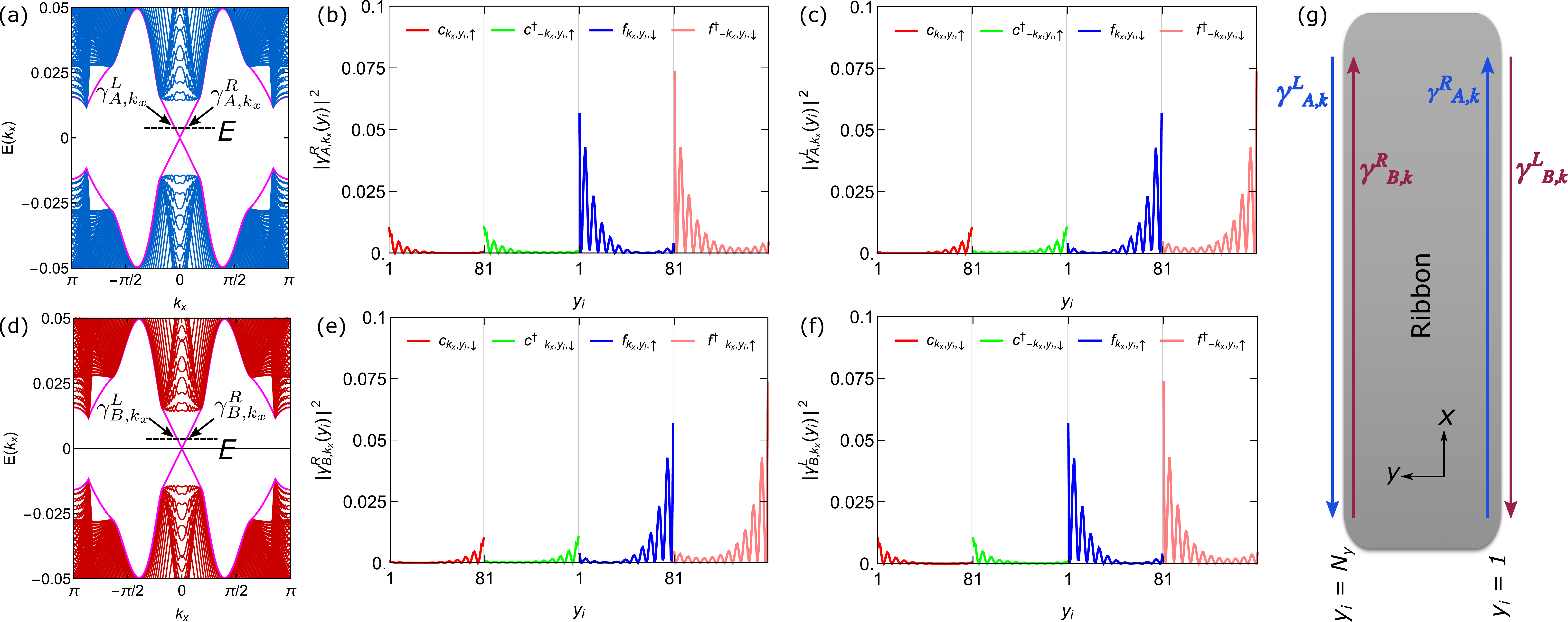}
    \caption{\added{Figures (a) and (d) show the Bogoliubov excitation spectra of $\mathcal{H}_A$ and $\mathcal{H}_B$,  respectively, near the chemical potential on a nano-strip with $N_y = 81$ chains. Figures (b), (c) and (e), (f) demonstrate the probability density of the Majorana edge state wave functions of $\mathcal{H}_A$ and $\mathcal{H}_B$ as a function of atom position $y_i$, $\left| \gamma^\Gamma_{I, k_x}(y_i)  \right|^2$  with $I = A,B$ and $\Gamma = R,L$ (pink curves in (a) and (d)), at a fixed energy $E \equiv E(k_x = \pm 0.03)$. The probability density is described by $\left| \gamma^\Gamma_{I, k_x} (y_i) \right|^2 = \left( \left| u^\Gamma_{I,k_x} \right|^2,\, \left| \bar{u}^\Gamma_{I,k_x} \right|^2,\, \left| v^\Gamma_{I,k_x} \right|^2, \, \left| \bar{v}^\Gamma_{I,k_x} \right|^2   \right) (y_i) $. The parameters are $J_K/t = 0.3$, $J_H/t = 1.0$, and doping $\delta = -0.3$. The edge states are of the helical type, as schematically represented in (g). }\deleted{The probability densities are calculated at finite momenta at $J_H/t = 0.3$. For $\mathcal{H}_{\mbd{k}}^{A}$ (top left in blue), $|\Psi_{A,k}|^2$ (top middle) dominates at one edge and $|\Psi_{A,-k}|^2$ (bottom middle) dominates at the other edge. And they reverse for $\mathcal{H}_{\mbd{k}}^{B}$ (bottom left in red and right panels). Clearly, $|\Psi_{A,k}|^2$ and $|\Psi_{B,-k}|^2$ have the same behavior and $|\Psi_{A,-k}|^2$ and $|\Psi_{B,k}|^2$ share the same behavior.}}
    \label{JH_3.0_HMF_AB}
\end{figure*}
\begin{figure*}[ht]
    \centering
    \includegraphics[width=0.9\textwidth]{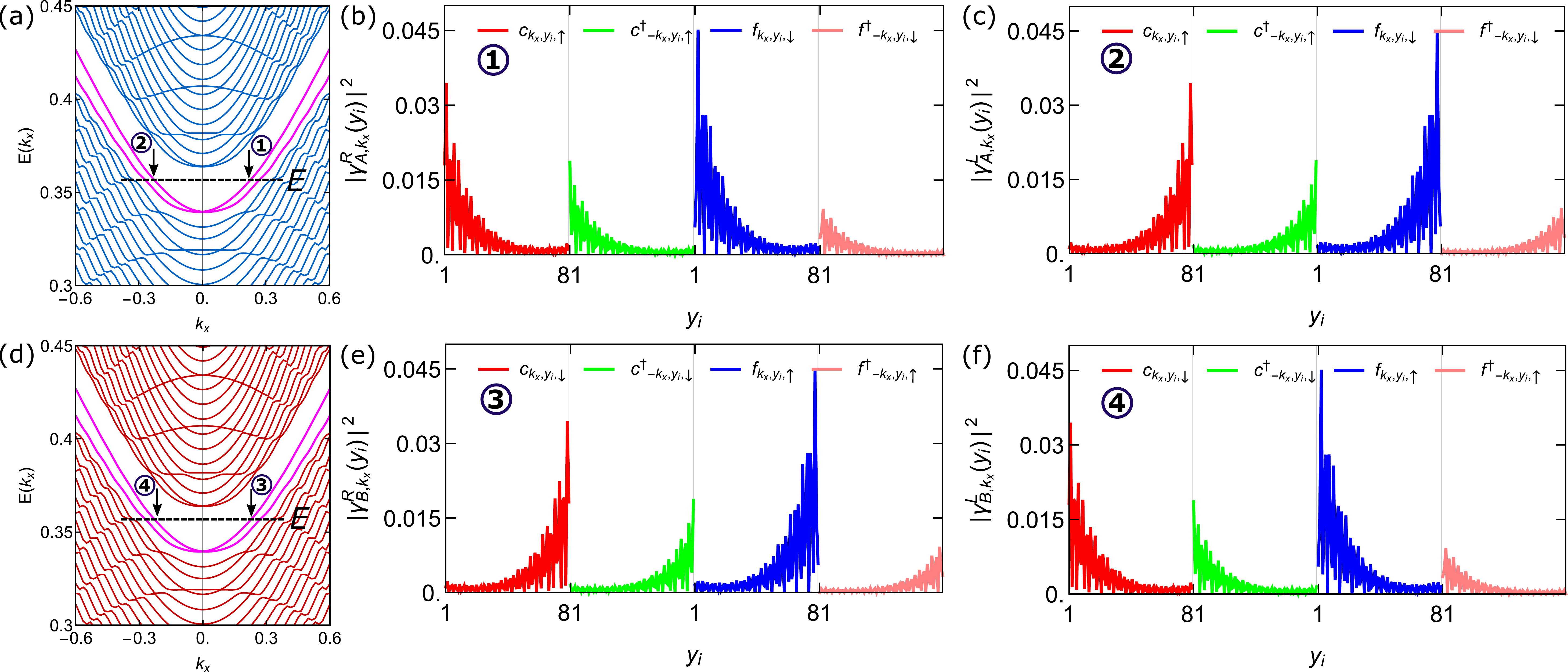}
    \caption{\added{The finite-energy ($E(k_x)>0$) Bogoliubov excitation spectra of (a) $\mathcal{H}_A$ (shown on top right in Fig. \ref{JH_3.0_HMF_AB}) and (d) $\mathcal{H}_B$. A pair of ``helical" edge states is found to exist at finite energy [pink curve in (a) and (d)], and their probability densities are shown in (b) and (c), (e) and (f), respectively, at a fixed energy $E(k_x = \pm 0.22)$.}}
    \label{fig:ribbon-upedgestate-Ny61}
\end{figure*}

\deleted{Via bulk-edge correspondence, the non-trivial topological $Z_2$ number $c_T$ of the bulk coexisting  superconducting state indicates a pair of gapless counter-propagating edge states to exist at edges. Accompanying with the particle-hole symmetry of our model, the edge state can therefore be regarded as helical Majorana fermion.  We confirm this expectation by examining the band structure and the edge-state wave function of the Kondo-RVB coexisting superconducting phase on a finite-sized strip}

We now check whether our model would support helical Majorana zero modes at the edge of a finite-sized system. We shall examine our model's band structures and edge-state wave functions on a finite-sized strip  that extends infinitely along the $x$ direction but contains a finite number of lattice sites in $y$.  The results are shown in Figs. \ref{JH_3.0_EBS_AB} to \ref{fig:ribbon-upedgestate-Ny61}. As shown in Fig. \ref{JH_3.0_EBS_AB},  gapless Dirac spectra of the Bogoliubov excitations around $k_x = 0$ near zero energy are observed, exhibiting one of the typical features of topological edge states. The excitations can be effectively described  by the linear-dispersed Hamiltonian $ \tilde{H}_{I}=\sum_{k_x}v_x|k_x| \left( \gamma^{R\,\dagger}_{I,k_x} \gamma^{R}_{I,k_x}-\gamma^{L\,\dagger}_{I,k_x}\gamma^L_{I,k_x}\right)$ with

\begin{align}
    \gamma_{I,k_x}^\Gamma =&\sum_{y_i}\left[u^\Gamma_{I,k_x}(y_i)c_{k_x,y_i,\uparrow}+\bar{u}^\Gamma_{I,k_x}(y_i) c_{-k_x,y_i,\uparrow}^{\dagger} \right.\nn
    &   \left.+v^\Gamma_{I,k_x}(y_i)f_{k_x,y_i,\downarrow}+\bar{v}^\Gamma_{I,k _x}(y_i)f_{-k_x,y_i,\downarrow}^{\dagger}\right]
    %\gamma_{A,k_x}^{L} & =  \sum_{y_i} \left[u_{A,k_x}(y_i)c_{-k_x,N_y+1-y_i,\uparrow} -\bar{u}_{A,k_x}(y_i) c_{k_x,N_y+1-y_i,\uparrow}^{\dagger} \right. \nn 
    %&\quad \,  \left. - v_{A,k_x}(y_i)f_{-k_x,N_y+1-y_i,\downarrow} 
    %+\bar{v}_{A,k_x}(y_i)f_{k_x,N_y+1-y_i,\downarrow}^{\dagger}\right],
    \label{eq:low-E-HA-gamma}
\end{align}

with $u,\bar{u}$ and $v,\bar{v}$ being the coherent factors. In Eq. (\ref{eq:low-E-HA-gamma}), $I \in A,B$, $\Gamma \in R,L $, and $\gamma^{R/L}_{A/B,k_x}$  represents the right/left-moving Bogoliubov quasiparticle  of $\tilde{H}_{A/B}$.  Here, $v_x$ in $\tilde{H}_I$ denotes the velocity. Due to time-reversal symmetry, $H_A$ is the time-reversal partner of $H_B$, and thus their   spectra are identical. \deleted{and the effective Hamiltonian of the low-energy excitations of $H_B$ takes a similar form to Eq. (\ref{eq:low-E-HA-gamma}).} The low-energy eigenstates with Dirac spectra near $k_x = 0$ for both $H_A$ and $H_B$ exhibit the typical property of edge states, as their probability densities accumulate mostly at the edges of strip, as shown in Fig. \ref{JH_3.0_HMF_AB}. Combining the directions of propagation inferred from the velocity $v_x \sim \partial E(k_x)/\partial k_x$, we can classify these edge states  into two groups, each of them constitutes a pair of counter-propagating  edge states (see Fig. \ref{JH_3.0_HMF_AB}), revealing the nature of helical Majorana zero modes. The helical type of the Majorana zero modes is the consequence of time-reversal symmetry of our model, reminiscent of the well-known Kane-Mele model on a single-layered graphene \cite{2005-PRL-KM-Z2,2005-PRL-KM}. Remarkably, in addition to the Majorana fermions at zero energy,  two pairs of  counter-propagating edge-states are observed at finite energy, see Fig. \ref{fig:ribbon-upedgestate-Ny61}. The two pairs of  edge states correspond to the edge states of the topological Kondo insulator, where the spin-triplet RVB order parameter is absent ($\Delta_t = 0 $) \cite{dzero-TKI-PRL,dzero-TKI-PRB,dzero-Ann-TKI}.

\begin{figure}
    \centering
    \includegraphics[width=0.45\textwidth]{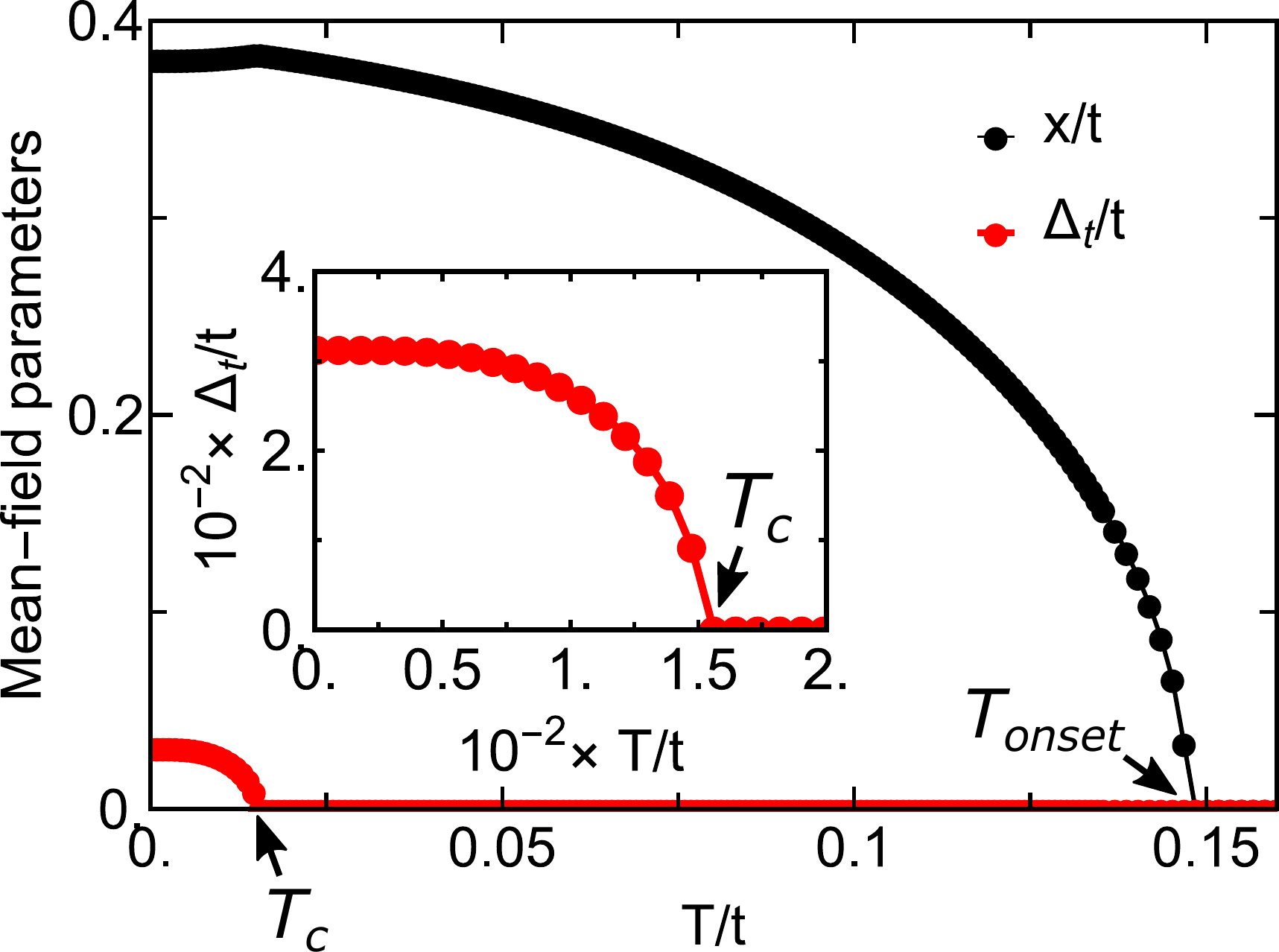}
    \caption{Plot of the temperature-dependent mean-field order parameters $x(T)/t$ and $\Delta_t (T)/t$ with $k_B = 1$, $J_K/t = 0.3$ and $J_H/t = 1.0$ fixed. Inset shows the enlarged plot of $\Delta_t (T)$. The single-impurity Kondo temperature occurs at $T_{\text{onset}}/t \approx 0.16$ while the transition of superconductivity takes places at temperature $T_c /t \approx 0.015$.}
    \label{fig:T-MF}
\end{figure}

%%%%%%%%%%%%%%%%%%%%%%%%%
%%%%%%%%%%%%%%%%%%%%%%%%%
\section{Discussions and Conclusions}

\added{We now discuss the application of our results for heavy-electron superconductors, particularly the Kondo lattice compound UTe$_2$. Experimental evidence indicates that this compound does not show  long-range magnetic order and is in the vicinity of the ferromagnetic quantum critical point, exhibiting both strong ferromagnetic fluctuations, possibly due to magnetic frustrations induced by sub-leading antiferromagnetic fluctuations \cite{YFYang-UTe, QM-UTe2-nature}, and Kondo screening \cite{Jiao-2020-UTe2,Ran-FM-UTe,Paglione-PRL-ARPES}. The DFT+$U$ calculations indicate that the dynamics of electron bands and the physical properties of UTe$_2$ are dominated by the electrons near the quasi-two-dimensional (cylindrical) Fermi surface with weak $k_z$ dependence despite its 3D crystal structure \cite{YFYang-UTe}. Superconductivity is reached at $T_c = 1.6$K, while the resistivity maximum observed at $T^\star \approx 15\sim 75$ $\text{K}$ reveals signature of coherent Kondo scattering \cite{rho-c-axis-PRB,Ran-FM-UTe}, indicating $T^\star/T_c \approx 10 \sim 50$. The superconductivity can, in general, co-exist and compete with the Kondo effect \cite{Jiao-2020-UTe2}. When a magnetic field is applied along the hard-magnetic axis $b$ of UTe$_2$ and before entering the superconducting phase,  a correlated paramagnetic phase is observed  below the temperature at which the magnetic susceptibility shows a broad maximum \cite{Braithwaite-CommPhys-PBphaseDiag}. Similar spin-liquid behavior has been observed in the magnetic susceptibility of another heavy fermion compound CePdAl  \cite{2019-Sun-CePdAl}. This similarity  suggests this correlated paramagnetic phase may feature  short-range magnetic order. Our theoretical framework based on  competition and collaboration between a Kondo-screened and a ferromagnetic $t$-RVB  spin-liquid states on a two-dimensional Kondo lattice is consistent with the above observations in UTe$_2$. It, therefore, constitutes a promising approach to account for its exotic phenomena. On the other hand, the chiral in-gap state, a signature of chiral topological superconductor, has been observed by scanning tunneling spectroscopy in the superconducting phase of UTe$_2$  \cite{Jiao-2020-UTe2}. Combining with the ferromagnetic fluctuations that are known to induce spin-triplet pairing, people believe UTe$_2$ is a promising candidate for the spin-triplet chiral  topological superconductor \cite{Jiao-2020-UTe2,Ran-FM-UTe}.  Furthermore, the superconducting phase co-existing with Kondo coherence in this material strongly suggests the role played by the Kondo effect in this possible topological superconductor. The topological Kondo superconducting state with equal-spin spin-triplet  $p$-wave   pairings we proposed here bears striking similarities to and strong relevance for the experimental observations on UTe$_2$: (i) the $d$- and $f$-orbitals electrons with their angular momentum quantum number  differing by $1$ in the uranium atoms of UTe$_2$ likely give rise to the odd-parity Kondo effect \cite{dzero-TKI-PRL,dzero-TKI-PRB,dzero-Ann-TKI}, (ii) the $t$-RVB state in our theory may be considered as one possible realization of the short-ranged ferromagnetic fluctuations in UTe$_2$, (iii) the Kondo-$t$-RVB co-existing superconducting state we find here qualitatively agrees with the co-existence between superconductivity and Kondo effect observed in UTe$_2$,  (iv) the high upper critical field exceeding the Pauli limit \cite{Ran-FM-UTe,Aoki-JPSJ-2020-FM} implies that the superconducting state of UTe$_2$ may have equal-spin Cooper pairs, and (v) the  effective pairing $\overline{\Delta}_{\bm{k}}^\sigma$ formed in the conduction band mentioned in Section \ref{sec:MFPD} shows  characteristics of  spin-triplet point-node gap structure \cite{point-node-Paglione}. Various characteristic temperature scales estimated from our mean-field calculations with $J_H/t = 1.0$ and $J_K/t = 0.3$ at finite temperatures agree reasonably well  with experimental observations (see Fig. \ref{fig:T-MF}): The superconducting transition temperature  $T_c$, theoretically determined from our mean-field analysis $T_c = \text{Min}[T(x = 0), T(\Delta_t = 0)]$, shows $T_c \approx 0.015 t \approx 2.3\text{\,K}$ by taking  estimated values of $t = 150\, \text{K}$ and  half-bandwidth  $D = 1.25t$ \cite{Paglione-PRL-ARPES}. The Kondo coherent scale can be obtained by $T^\star = x^2(T=0)/D \approx 17.4\,\text{K}$ \cite{Burdin-PRL}. The  ratio $T^\star/T_c \approx 8$ is in reasonable agreement with experimental observations.  The onset temperature $T_{\text{onset}}$ of  Kondo hybridization, which occurs at $x(T=T_{\text{onset}}) = 0$, displays $T_{\text{onset}} \approx 0.16t \approx 24\,\text{K}$, within the theoretically estimated range $10\text{K}< T_{\text{onset}}<100 \text{K}$ by DMFT calculation \cite{Paglione-PRL-ARPES}. Meanwhile, there have been evidences of TRS breaking in UTe$_2$ from the observed two superconducting transitions and a finite polar Kerr effect at $T<T_c$ \cite{Hayes-KerrEff-Science}, likely due to proximity to the ferromagnetic ordered phase.  A number of theoretical attempts were proposed based on these observations \cite{Agterberg-topo-band-PRB,Coleman-2022-UTe}. However, the observed single superconducting transition near ambient pressure and zero field \cite{Braithwaite-CommPhys-PBphaseDiag,  Rosa-SingleTran-CommMat, Rousel-arxiv-singletransition} as well as the theoretically proposed unitary triplet pairing \cite{YFYang-UTe} suggest TRS may be preserved in UTe$_2$. Though our results shown above are obtained in the presence of TRS, the chiral $p$-wave superconducting state with chiral Majorana zero mode at edges  is expected to occur here once a  time-reversal breaking magnetic field is applied \cite{sato-PRB-NSC}. Our distinct predictions with and without fields serve as theoretical guidance for future experiments to distinguish the time-reversal breaking from time-reversal preserving triplet pairing states  in UTe$_2$.  Since the Kondo correlations stabilize the $t$-RVB spin liquid in the co-existing superconducting phase, it is expected to be robust against gauge-field fluctuations beyond the mean field.  Our approach and results are distinct from the spin-triplet non-topological superconducting state recently proposed based on the Hund's-Kondo coupling and $S_z=0$ $t$-RVB state to account for UTe$_2$ \cite{Coleman-2022-UTe}.} 

%we have performed a similar analysis in Supp. ? by including a magnetic field to explicitly break TRS. Therein, the TRS $p \pm i p^\prime$ paired superconducting state with helical Majorana zero modes at edges reduces to TRS breaking chiral $p+i p^\prime$-wave superconducting state with chiral Majorana zero modes.}

In conclusion, we propose \replaced{a first  realization of}{a microscopic mechanism} the topological superconductivity in the Kondo lattice model, \added{a distinct class of topological superconductors due to purely  strong electron correlations without employing spin-orbit coupling or  proximity effect. } \deleted{which is relevant for possible topological heavy-fermion superconducting states in heavy-fermion compounds.} \added{A topological Kondo superconductor essentially constitutes of 1)  itinerant $c$ and  localized $f$ bands with different orbital quantum numbers, 2) strong Hubbard interaction of the $f$ electrons, 3) odd-parity Kondo hybridization of the $c$ and $f$ bands, and 4) the attractive exchange interaction of the $f$ electrons with spin-triplet correlations.} Starting from the odd-parity Anderson lattice model, we obtain the unconventional type of Kondo hybridization and \added{ferromagnetic RKKY-like} interaction via perturbation theory, \added{leading} \deleted{which lead} to spin-triplet resonating-valence-bond (RVB) pairing between $f$-electrons with \added{time-reversal invariant} $p \pm ip^\prime$-wave gap symmetry. Via the mean-field approach, we find a Kondo triplet-RVB coexisting phase in the intermediate range of the Kondo to RKKY coupling ratio. This phase is shown as a time-reversal invariant topological superconducting state with a spin-triplet $p \pm ip^\prime$-wave RVB pairing gap. It exhibits non-trivial topology in the bulk band structure, and supports helical Majorana zero modes at edges. \added{Our prediction in the presence of a time-reversal breaking field leads to chiral $p$-wave spin-triplet topological  Kondo superconductor.  Our results on the superconducting transition temperature, Kondo coherent scale, and onset temperature of Kondo hybridization  not only qualitatively but also  quantitatively agree with the observations for UTe$_2$. The theoretical framework we propose here opens up the search for topological superconductors induced by strongly electronic correlations on the Kondo lattice compounds.  
%a qualitative description of heavy-fermion Kondo lattice UTe$_2$, a strong candidate for the chiral $p$-wave topological Kondo superconductor.
} 

\section{Acknowledgements} This work is supported by the Ministry of Science and Technology Grants 104-2112-M-009-004-MY3 and 107-2112-M-009-010-MY3, the National Center for Theoretical Sciences of Taiwan, Republic of China (to C.-H. C.).
%%%%%%%%%%%%%%%%%%%%%%%%%
%%%%%%%%%%%%%%%%%%%%%%%%%
\appendix
\section{The Schrieffer-Wolff  transformation (SWT)}
\label{app:SWT}
In this section, we provide derivations of the Kondo term via using the SWT.  We first perform the SWT on an odd-parity single-impurity Anderson model where an impurity at an arbitrary site $i$ hybridizes with the conduction electrons on the four nearest-neighbor sites of $i$. This result will be successively  generalized to the lattice version. 

The single-impurity Anderson model takes the following form
%\begin{widetext}
\begin{align}
    H=&\sum_{\mbd{k}\sigma}\varepsilon_{\mbd{k}}c_{\mbd{k}\sigma}^{\dagger}c_{{\mbd{k}}\sigma}+\sum_{\sigma}\varepsilon_{f}f_{i\sigma}^{\dagger}f_{i\sigma} +Un_{i\uparrow}^{f}n_{i\downarrow}^{f} \nn
    & +\sum_{\sigma\sigma^{\prime}}\sum_{\alpha=x,y}\left[iV\nu_{\hat{\alpha}}\sigma_{\alpha}^{\sigma\sigma^{\prime}}c_{i+\hat{\alpha},\sigma}^{\dagger}f_{i\sigma^{\prime}} +H.c. \right],
  %  & \qquad \qquad \left. +iV\nu_{\hat{\alpha}}\sigma_{\alpha}^{\sigma\sigma^{\prime}}f_{i\sigma}^{\dagger}c_{i-\hat{\alpha},\sigma^{\prime}}\right],
    \label{eq:H-SAM}
\end{align}
%\end{widetext}
where $\hat{\alpha}\equiv\pm\hat{x},\pm\hat{y}$ denotes the nearest-neighbor vectors of a  square lattice, and $\nu _{\hat{\alpha}}$ satisfies $\nu _{\hat{\alpha}}=-\nu_{-\hat{\alpha}}$ and $\nu_{\hat{x}}=\nu_{\hat{y}}=1$. 

The SWT aims at projecting out the empty and doubly occupied states to generate the effective Hamiltonian $H_{eff}$ in the Kondo (singly-occupied) limit. Following Ref. \cite{hewson1997kondo}, we first use the states of impurity occupation as the basis set, $\lbrace | f^0 \rangle, \,\, | f^1 \rangle, \,\, | f^2 \rangle  \rbrace$ with the superscripts  being denoted as the occupation of the localized electrons, to expand the Hamiltonian of Eq. (\ref{eq:H-SAM}) in the following matrix form,
\begin{align}
    H= \left[\begin{array}{ccc}
H_{00} & H_{01} & H_{02}\\
H_{10} & H_{11} & H_{12}\\
H_{20} & H_{21} & H_{22}
\end{array}\right].
\label{eq:H-SWT}
\end{align}
The matrix elements of Eq. (\ref{eq:H-SWT}), denoted as $H_{ij} \equiv \langle f^i | H | f^j \rangle $ with $i,j = 0,1,2$, are
    \begin{align}
        H_{10}&=\sum_{\sigma\sigma^{\prime}}\sum_{\alpha=\pm x,\pm y}iV\nu_{\hat{\alpha}}\sigma_{\alpha}^{\sigma\sigma^{\prime}}f_{i\sigma}^{\dagger}c_{i-\hat{\alpha},\sigma^{\prime}}
        =H_{21}, \nn [6pt]
        H_{01}&=H_{10}^{\dagger}=\sum_{\sigma\sigma^{\prime}}\sum_{\alpha=\pm x, \pm y}iV\nu_{\hat{\alpha}}\sigma_{\alpha}^{\sigma\sigma^{\prime}}c_{i+\hat{\alpha},\sigma}^{\dagger}f_{i\sigma^{\prime}}=H_{12}, \nn [6pt]
        H_{11}& = \sum_{\mbd{k}\sigma}\varepsilon_{\mbd{k}}c_{\mbd{k}\sigma}^{\dagger}c_{\mbd{k}\sigma}+\sum_{\sigma}\varepsilon_{f}f_{i\sigma}^{\dagger}f_{i\sigma},\,\, H_{00}=\sum_{\mbd{k}\sigma}\varepsilon_{\mbd{k}}c_{\mbd{k}\sigma}^{\dagger}c_{\mbd{k}\sigma},\nn[5pt] H_{22}&=\sum_{\mbd{k}\sigma}\varepsilon_{\mbd{k}}c_{\mbd{k}\sigma}^{\dagger}c_{\mbd{k}\sigma}+\sum_{\sigma}\varepsilon_{f}f_{j\sigma}^{\dagger}f_{j\sigma}+Un_{i\uparrow}^{f}n_{i\downarrow}^{f}.
    \end{align}
We then project out $|f^0 \rangle$ and $|f^2 \rangle$ from the Hilbert space to obtain the effective Hamiltonian $H_{eff}$ at the  Kondo limit satisfying  $H_{eff} | f^1 \rangle =E | f^1 \rangle$ with $E$ being the eigenenergy. Via Eq. (\ref{eq:H-SWT}), $H_{eff}$ can be expressed as $H_{eff} =  H_{11} + H^\prime$ with
 \begin{widetext}
    \begin{align}
        H^\prime = &  H_{10}(E-H_{00})^{-1}H_{01} 
        + H_{12}(E-H_{22})^{-1}H_{21} \nn
             = &\sum_{\alpha,\alpha^\prime=x,y}\sum_{\sigma\sigma^{\prime}}\sum_{\sigma^{\prime\prime}\sigma^{\prime\prime\prime}}\left[\frac{V^2}{\varepsilon_{F}-\varepsilon_{f}-U}\left(i\nu_{\hat{\alpha}}\sigma_{\alpha}^{\sigma\sigma^{\prime}}c_{i+\hat{\alpha},\sigma}^{\dagger}f_{i\sigma^{\prime}}\right)\left(i\nu_{\hat{\alpha}^{\prime}}\sigma_{\alpha^{\prime}}^{\sigma^{\prime\prime}\sigma^{\prime\prime\prime}}f_{i\sigma^{\prime\prime}}^{\dagger}c_{i-\hat{\alpha}^{\prime},\sigma^{\prime\prime\prime}}\right) \right.\\
        &\left. \qquad\qquad +\frac{V^2}{\varepsilon_{f}-\varepsilon_{F}}\left(i\nu_{\hat{\alpha}}\sigma_{\alpha}^{\sigma\sigma^{\prime}}f_{i\sigma}^{\dagger}c_{i-\hat{\alpha},\sigma^{\prime}}\right)\left(i\nu_{\hat{\alpha}^{\prime}}\sigma_{\alpha^{\prime}}^{\sigma^{\prime\prime}\sigma^{\prime\prime\prime}}c_{i+\hat{\alpha}^{\prime},\sigma^{\prime\prime}}^{\dagger}f_{i\sigma^{\prime\prime\prime}}\right) \right]
        \label{eq:H-prime}
    \end{align}
\end{widetext}
Here, we skip the derivations of $H_{10}(E-H_{00})^{-1}H_{01}$ and $H_{12}(E-H_{22})^{-1}H_{21}$ in Eq. (\ref{eq:H-prime}) as those are standard and can be found in a number of references. See, for example, Refs. \cite{SW-transformation,hewson1997kondo}. $H^\prime$ can be further cast into the form similar to the conventional single-impurity Kondo term, with the following antiferromagnetic Kondo coupling
\begin{align}
    J_{K}=\frac{V^{2}}{U+\varepsilon_{f}-\varepsilon_{F}}+\frac{V^{2}}{\varepsilon_{F}-\varepsilon_{f}}>0,
\end{align}
plus a potential scattering term. Eq. (\ref{eq:H-prime}) can be generalized to the lattice version by summing over all lattice sites, as described by Eq. (\ref{eq:HK}). 

\section{Derivation of the effective ferromagnetic RKKY-like interaction}
\label{app:rkky}
In the section, we derive  the RKKY-like interaction by perturbatively expanding   $H_K$ of Eq. (\ref{eq:HK}) to second order. 

The unperturbed state is described as
\begin{align}
    \left|0,f\right\rangle=\left|k_{1}m_{1},k_{2}m_{2},\cdots,k_{N}m_{N}\right\rangle \left|f\right\rangle,
    \label{eq:unper-state}
\end{align}
where conduction electrons do not interact with the impurities. In Eq. (\ref{eq:unper-state}),  $\left|k_{1}m_{1},k_{2}m_{2},\cdots,k_{N}m_{N}\right\rangle$ represents the Fermi sea with all wave vectors lying below the Fermi wave vector, namely  $k_{i}<k_{F}$. After imposing perturbation, the unperturbed state acquires correction and the  corrected eigenenergy is expressed in powers of $J_K$, $E=E_0 +\Delta E^{(1)}+\Delta E^{(2)}+O(J_K^3)$ with $E_0$ being the eigenenergy of the unperturbed state. 

The first and second order energy corrections take the form
\begin{align}
    &\Delta E^{(1)}=\left\langle0,f\right|H_{K}\left|0,f\right\rangle, \nn
    &\Delta E^{(2)}=\sum_{(0,f)\neq(A,f')}\frac{\left|\left\langle0,f\right|H_{K}\left|A,f^{\prime}\right\rangle\right|^{2}}{E_{0}-E_{A}},
    \label{eq:E-correction}
\end{align}
 where $| A,f^\prime \rangle$ denotes the excited state which can be expressed as a direct product of the building blocks $|k_{i}^{\prime\prime},m_{i}^{\prime\prime}\rangle$, with part of wave vectors lying above the Fermi surface, i.e. $k_{i}^{\prime\prime}>k_{F}$.
 
Here, we first derive the effective interaction of the $f$ fermions for a simpler two-impurity model and generalize the results to the lattice version.
%For simplicity, we first assume tht the system of our interest contains only two impurities, and successively evaluate $\Delta E^{(1)}$ and $\Delta E^{(2)}$ under perturbation due to $H_K$. We then generalize the results of two-impurity case to its lattice version. 

$\Delta E^{(1)}$ can be evaluated by summing over the subspace of the conduction electron, yielding
\begin{align}
    \Delta E^{(1)} & = \left\langle0,f\right|H_{K}\left|0,f\right\rangle \nn & = \frac{4 n_f J_{K}}{\mathcal{N}_{s}}\sum_{\bm{k}<k_F}\left(
\sin^{2}k_{x}+\sin^{2}k_{y}\right)+\mathcal{C},
    \label{eq:Delta-E-1}
\end{align}
where $n_{f}=\sum_{i=1,2,\sigma} \left\langle f \right|f_{i\sigma}^{\dagger}f_{i\sigma} \left| f\right\rangle$ and $\mathcal{C}$ is a constant.  $H_K$ in Eq. (\ref{eq:Delta-E-1}) denotes the two-impurity Kondo term.  It turns out that $\Delta E^{(1)}$ only introduces a constant energy shift for the bare energy level of the $f$ fermions. 

$\Delta E^{(2)}$ is given by 
\begin{widetext}
\begin{align}
   \Delta E^{(2)} %&\frac{1}{\mathcal{N}_{s}^{4}}\sum_{f^{\prime\prime}}\sum_{k_{1}^{\prime\prime}}\sum_{m_{1}^{\prime\prime}}\cdots\sum_{k_{N}^{\prime\prime}}\sum_{m_{N}^{\prime\prime}}\left(\frac{J_{K}^{2}}{E_{0}-E_{A}}\right)\sum_{i=1}^{2}\sum_{\sigma\sigma^{\prime}}\sum_{\sigma^{\prime\prime}\sigma^{\prime\prime\prime}}\sum_{\alpha,\alpha^{\prime}}\sum_{\mbd{k},\mbd{k}^{\prime}}\left(\sum_{l=1}^{2}\sum_{\mu,\mu^{\prime}}\sum_{\tau\tau^{\prime}}\sum_{\tau^{\prime\prime}\tau^{\prime\prime\prime}}\sum_{\beta,\beta^{\prime}}\sum_{\mbd{q},\mbd{q}^{\prime}}\right)
  % \nn 
   %& \quad \left[e^{i\mbd{k}\cdot(\mbd{r}_{i}+\hat{\alpha})-i\mbd{k}^{\prime}\cdot(\mbd{r}_{i}-\hat{\alpha}^{\prime})}i\nu_{\hat{\alpha}}\sigma_{\alpha}^{\sigma\sigma^{\prime}}i\nu_{\hat{\alpha}^{\prime}}\sigma_{\alpha^{\prime}}^{\sigma^{\prime\prime}\sigma^{\prime\prime\prime}}\right]e^{i\mbd{q}\cdot(\mbd{r}_{l}+\hat{\beta})-i\mbd{q}^{\prime}\cdot(\mbd{r}_{l}-\hat{\beta}^{\prime})}\left(i\nu_{\hat{\beta}}\sigma_{\beta}^{\tau\tau^{\prime}}i\nu_{\hat{\beta}^{\prime}}\sigma_{\beta^{\prime}}^{\tau^{\prime\prime}\tau^{\prime\prime\prime}}\right)
   %\nn
   %& \quad \langle f\Big|\left\langle k_{1}m_{1},k_{2}m_{2},\cdots,k_{N}m_{N}\right|\left(c_{\mbd{k}\sigma}^{\dagger}f_{i\sigma^{\prime}}f_{i\sigma^{\prime\prime}}^{\dagger}c_{\mbd{k}^{\prime}\sigma^{\prime\prime\prime}}\right)\left|k_{1}^{\prime\prime}m_{1}^{\prime\prime},\,k_{2}^{\prime\prime}m_{2}^{\prime\prime},\cdots,k_{N}^{\prime\prime}m_{N}^{\prime\prime}\right\rangle \left|f^{\prime\prime}\right\rangle 
   %\nn 
   %& \quad\langle f^{\prime\prime}\Big|\left\langle k_{1}^{\prime\prime}m_{1}^{\prime\prime},\,k_{2}^{\prime\prime}m_{2}^{\prime\prime},\cdots,k_{N}^{\prime\prime}m_{N}^{\prime\prime}\right|\left(c_{\mbd{q}\tau}^{\dagger}f_{l\tau^{\prime}}f_{l\tau^{\prime\prime}}^{\dagger}c_{\mbd{q}^{\prime}\tau^{\prime\prime\prime}}\right)\left|k_{1}m_{1},k_{2}m_{2},\cdots,k_{N}m_{N}\right\rangle \left|f\right\rangle 
   =&\frac{1}{\mathcal{N}_{s}^{4}}\sum_{f^{\prime\prime}}\sum_{k_{1}^{\prime\prime}}\sum_{m_{1}^{\prime\prime}}\cdots\sum_{k_{N}^{\prime\prime}}\sum_{m_{N}^{\prime\prime}}\left(\frac{J_{K}^{2}}{E_{0}-E_{A}}\right)\sum_{i=1}^{2}\sum_{\sigma\sigma^{\prime}}\sum_{\sigma^{\prime\prime}\sigma^{\prime\prime\prime}}\sum_{\alpha,\alpha^{\prime}}\sum_{\mbd{k},\mbd{k}^{\prime}}\left(\sum_{j=1}^{2}\sum_{\tau\tau^{\prime}}\sum_{\tau^{\prime\prime}\tau^{\prime\prime\prime}}\sum_{\beta,\beta^{\prime}}\sum_{\mbd{q},\mbd{q}^{\prime}}\right)
   \nn 
   &\quad \times \left[e^{i\mbd{k}\cdot(\mbd{r}_{i}+\hat{\alpha})-i\mbd{k}^{\prime}\cdot(\mbd{r}_{i}-\hat{\alpha}^\prime)}i\nu_{\hat{\alpha}}i\nu_{\hat{\alpha}^\prime}\right]\left[e^{i\mbd{q}\cdot(\mbd{r}_{j}+\hat{\beta})-i\mbd{q}^{\prime}\cdot(\mbd{r}_{j}-\hat{\beta}^\prime)}i\nu_{\hat{\beta}}i\nu_{\hat{\beta}^\prime}\right]
   \nn  
  & \quad \times \sigma_{\alpha}^{\sigma\sigma^{\prime}}\sigma_{\alpha^{\prime}}^{\sigma^{\prime\prime}\sigma^{\prime\prime\prime}}\left\langle f\right|f_{i\sigma^{\prime}}f_{i\sigma^{\prime\prime}}^{\dagger}\left|f^{\prime\prime}\right\rangle \left\langle k_{1}m_{1},k_{2}m_{2},\cdots,k_{N}m_{N}\right|\left(c_{\mbd{k}\sigma}^{\dagger}c_{\mbd{k}^{\prime}\sigma^{\prime\prime\prime}}\right)\left|k_{1}^{\prime\prime}m_{1}^{\prime\prime},\,k_{2}^{\prime\prime}m_{2}^{\prime\prime},\cdots,k_{N}^{\prime\prime}m_{N}^{\prime\prime}\right\rangle 
   \nn
   & \quad \times \sigma_{\beta}^{\tau\tau^{\prime}}\sigma_{\beta^{\prime}}^{\tau^{\prime\prime}\tau^{\prime\prime\prime}}\left\langle f^{\prime\prime}\right|f_{j\tau^{\prime}}f_{j\tau^{\prime\prime}}^{\dagger}\left|f\right\rangle \left\langle k_{1}^{\prime\prime}m_{1}^{\prime\prime},\,k_{2}^{\prime\prime}m_{2}^{\prime\prime},\cdots,k_{N}^{\prime\prime}m_{N}^{\prime\prime}\right|\left(c_{\mbd{q}\tau}^{\dagger}c_{\mbd{q}^{\prime}\tau^{\prime\prime\prime}}\right)\left|k_{1}m_{1},k_{2}m_{2},\cdots,k_{N}m_{N}\right\rangle. 
\end{align}
\end{widetext}

An annihilation operator acts on $\left|A,f^{\prime}\right\rangle$  can be obtained as 

\begin{align}
    & c_{q\sigma^{\prime}} \left|k_{1}^{\prime\prime}m_{1}^{\prime\prime},\,k_{2}^{\prime\prime}m_{2}^{\prime\prime},\cdots,k_{N}^{\prime\prime}m_{N}^{\prime\prime}\right\rangle 
    \nn
     = &\sum_{\alpha=1}^{N}(-1)^{p_{\alpha}}\delta_{q,k_{\alpha}^{\prime\prime}}c_{\sigma^{\prime}}\left|\left(\prod_{l=1}^{\alpha-1}k_{l}^{\prime\prime}m_{l}^{\prime\prime}\right)m_{\alpha}^{\prime\prime}\left(\prod_{l=\alpha+1}^{N}k_{l}^{\prime\prime}m_{l}^{\prime\prime}\right)\right\rangle, 
\end{align}

we can thus obtain

\begin{widetext}
\begin{align}
    &\left\langle k_{1}m_{1},k_{2}m_{2},\cdots,k_{N}m_{N}\right|\left(c_{\mbd{k}\sigma}^{\dagger}c_{\mbd{k}^{\prime}\sigma^{\prime\prime\prime}}\right)\left|k_{1}^{\prime\prime}m_{1}^{\prime\prime},\,k_{2}^{\prime\prime}m_{2}^{\prime\prime},\cdots,k_{N}^{\prime\prime}m_{N}^{\prime\prime}\right\rangle 
    \nn 
    =&\sum_{\alpha=1}^{N}\sum_{\beta=1}^{N}(-1)^{p_{\alpha}}(-1)^{p_{\beta}}\delta_{\mbd{k},\mbd{k}_{\alpha}}\delta_{\mbd{k}^{\prime},\mbd{k}_{\beta}^{\prime\prime}}\left\langle m_{\alpha}\right|c_{\sigma}^{\dagger}c_{\sigma^{\prime\prime\prime}}\left|m_{\beta}^{\prime\prime}\right\rangle
    \left\langle \left(\prod_{l=1}^{\alpha-1}k_{l}m_{l}\right)\left(\prod_{l=\alpha+1}^{N}k_{l}m_{l}\right)\right|\left(\prod_{l=1}^{\beta-1}k_{l}^{\prime\prime}m_{l}^{\prime\prime}\right)\left(\prod_{l=\beta+1}^{N}k_{l}^{\prime\prime}m_{l}^{\prime\prime}\right)\Bigg>.
\end{align}
\end{widetext}
The above matrix element is nonzero only if the momentum is restricted by certain constraints and $(k_{i},m_{i})=(k_{i}^{\prime\prime},m_{i}^{\prime\prime})$, signifying $p_{\alpha}=p_{\beta}$:
%\begin{widetext}
\begin{align}
    &\left\langle k_{1}m_{1},\cdots,k_{N}m_{N}\right|\left(c_{\mbd{k}\sigma}^{\dagger}c_{\mbd{k}^{\prime}\sigma^{\prime\prime\prime}}\right)\left|k_{1}^{\prime\prime}m_{1}^{\prime\prime},\cdots,k_{N}^{\prime\prime}m_{N}^{\prime\prime}\right\rangle \nn
    %= &\Theta(k_{F}-\left|\mbd{k}\right|)\Theta\left(\left|\mbd{k}^{\prime}\right|-k_{F}\right)\sum_{\alpha=1}^{N}\delta_{\mbd{k},\mbd{k}_{\alpha}}\delta_{\mbd{k}^{\prime},\mbd{k}_{\alpha}^{\prime\prime}}\langle m_{\alpha}\Big|c_{\sigma}^{\dagger}c_{\sigma^{\prime\prime\prime}}\Big|m_{\alpha}^{\prime\prime}\rangle
   % \nn
    %&\qquad\qquad\left\langle \left(\Pi_{j^{\prime}=1}^{\alpha-1}k_{j^{\prime}}m_{j^{\prime}}\right)\left(\Pi_{j^{\prime}=\alpha+1}^{N}k_{j^{\prime}}m_{j^{\prime}}\right)\right|\left(\Pi_{j^{\prime}=1}^{\alpha-1}k_{j^{\prime}}^{\prime\prime}m_{j^{\prime}}^{\prime\prime}\right)\left(\Pi_{j^{\prime}=\alpha+1}^{N}k_{j^{\prime}}^{\prime\prime}m_{j^{\prime}}^{\prime\prime}\right)\Big>
    %\nn
   = & \Theta(k_{F}-\left|\mbd{k}\right|)\Theta\left(\left|\mbd{k}^{\prime}\right|-k_{F}\right)
   \nn
   &\times \sum_{\alpha=1}^{N}\left[ \delta_{\mbd{k},\mbd{k}_{\alpha}}\delta_{\mbd{k}^{\prime},\mbd{k}_{\alpha}^{\prime\prime}}  \left\langle m_{\alpha}\right|c_{\sigma}^{\dagger}c_{\sigma^{\prime\prime\prime}}\left|m_{\alpha}^{\prime\prime}\right\rangle \prod_{l\neq\alpha}\delta_{\mbd{k}_{l}\mbd{k}_{l}^{\prime\prime}}\delta_{m_{l}m_{l}^{\prime\prime}}\right]
\end{align}
%\end{widetext}
Plugging this into $\Delta E^{(2)}$, we have
\begin{widetext}
\begin{align}
    \Delta E^{(2)}=&\frac{1}{\mathcal{N}_{s}^{4}}\sum_{f^{\prime\prime}}\sum_{a=1}^{N}\sum_{k_{a}^{\prime\prime}}\sum_{m_{a}^{\prime\prime}}\left(\frac{J_{K}^{2}}{E_{0}-E_{A}}\right)\sum_{i=1}^{2}\sum_{\sigma\sigma^{\prime}}\sum_{\sigma^{\prime\prime}\sigma^{\prime\prime\prime}}\sum_{\alpha,\alpha^{\prime}}\left(\sum_{j=1}^{2}\sum_{\tau\tau^{\prime}}\sum_{\tau^{\prime\prime}\tau^{\prime\prime\prime}}\sum_{\beta,\beta^{\prime}}\sum_{\mbd{q},\mbd{q}^{\prime}}\right)
    \nn
    &\qquad \times \Theta(k_{F}-\left|\mbd{k}_{a}\right|)\Theta\left(\left|\mbd{k}_{a}^{\prime\prime}\right|-k_{F}\right)\left[e^{i\mbd{k}_{a}\cdot(\mbd{r}_{i}+\hat{\alpha})-i\mbd{k}_{a}^{\prime\prime}\cdot(\mbd{r}_{i}-\hat{\alpha}^{\prime})}i\nu_{\hat{\alpha}}i\nu_{\hat{\alpha}^{\prime}}\right]\left[e^{i\mbd{q}\cdot(\mbd{r}_{j}+\hat{\beta})-i\mbd{q}^{\prime}\cdot(\mbd{r}_{j}-\hat{\beta}^{\prime})}i\nu_{\hat{\beta}}i\nu_{\hat{\beta}^{\prime}}\right]\nn
   & \qquad\times \sigma_{\alpha}^{\sigma\sigma^{\prime}}\sigma_{\alpha^{\prime}}^{\sigma^{\prime\prime}\sigma^{\prime\prime\prime}}\sigma_{\beta}^{\tau\tau^{\prime}}\sigma_{\beta^{\prime}}^{\tau^{\prime\prime}\tau^{\prime\prime\prime}}\left\langle f\right|f_{i\sigma^{\prime}}f_{i\sigma^{\prime\prime}}^{\dagger}\left|f^{\prime\prime}\right\rangle \left\langle f^{\prime\prime}\right|f_{j\tau^{\prime}}f_{j\tau^{\prime\prime}}^{\dagger}\left|f\right\rangle \left\langle m_{a}\right|c_{\sigma}^{\dagger}c_{\sigma^{\prime\prime\prime}}\left|m_{a}^{\prime\prime}\right\rangle
    \nn
    & \qquad\times \left\langle \left(\prod_{l=1}^{a-1}k_{l}m_{l}\right)k_{a}^{\prime\prime}m_{a}^{\prime\prime}\left(\prod_{l=a+1}^{N}k_{l}m_{l}\right)\right|\left(c_{\mbd{q}\tau}^{\dagger}c_{\mbd{q}^{\prime}\tau^{\prime\prime\prime}}\right)\left|k_{1}m_{1},k_{2}m_{2},\cdots,k_{N}m_{N}\right\rangle 
    \label{eq:E(2)-2}
\end{align}
\end{widetext}
The matrix element of $c^\dagger_{\mbd{q}\tau}c_{\mbd{q}^\prime \tau^{\prime\prime\prime}}$ in the fourth line of Eq. (\ref{eq:E(2)-2}) can be evaluated as 
\begin{widetext}
\begin{align}
    \left\langle \left(\prod_{l=1}^{a-1}k_{l}m_{l}\right)k_{a}^{\prime\prime}m_{a}^{\prime\prime}\left(\prod_{l=a+1}^{N}k_{l}m_{l}\right)\right|\left(c_{\mbd{q}\tau}^{\dagger}c_{\mbd{q}^{\prime}\tau^{\prime\prime\prime}}\right)\left|k_{1}m_{1},k_{2}m_{2},\cdots,k_{N}m_{N}\right\rangle = \delta_{\mbd{q},\mbd{k}_{a}^{\prime\prime}}\delta_{\mbd{q}^{\prime},\mbd{k}_{\alpha}}\langle m_{a}^{\prime\prime}|c_{\tau}^{\dagger}c_{\tau^{\prime\prime\prime}}|m_{a}\rangle.
\end{align}
\end{widetext}
Hence, the energy correction $\Delta E^{(2)}$ can be further simplified as (sum over $f^{\prime\prime},m_{a}^{\prime\prime},\mbd{q},\mbd{q}^{\prime}$ and suppress the subscript $a$ below)
\begin{widetext}
\begin{align}
    \Delta E^{(2)}=&\frac{1}{\mathcal{N}_{s}^{4}}\sum_{i,j=1}^{2}\sum_{\alpha,\alpha^{\prime}}\sum_{\varepsilon_\mbd{k}<\mu}\sum_{\varepsilon_{\mbd{k}^{\prime\prime}} >\mu}\sum_{m,\tau=\pm}\sum_{\beta,\beta^{\prime}}\left(\frac{J_{K}^{2}}{\varepsilon_{\bm{k}}-\varepsilon_{\bm{k}^{\prime\prime}}}\right)\left(i\nu_{\hat{\alpha}}i\nu_{\hat{\alpha}^{\prime}}i\nu_{\hat{\beta}}i\nu_{\hat{\beta}^{\prime}}\right)\nn[5pt]
    &\quad \times e^{i\mbd{k}\cdot(\mbd{r}_{i}+\hat{\alpha})-i\mbd{k}^{\prime\prime}\cdot(\mbd{r}_{i}-\hat{\alpha}^{\prime})}e^{i\mbd{k}^{\prime\prime}\cdot(\mbd{r}_{j}+\hat{\beta})-i\mbd{k}\cdot(\mbd{r}_{j}-\hat{\beta}^{\prime})}\sigma_{\alpha}^{m,-m}\sigma_{\alpha^{\prime}}^{-\tau,\tau}\sigma_{\beta}^{\tau,-\tau}\sigma_{\beta^{\prime}}^{-m,m}\left\langle f\right|f_{i,-m}f_{i,-\tau}^{\dagger}f_{j,-\tau}f_{j,-m}^{\dagger}\left|f\right\rangle 
\end{align}
\end{widetext}
The effective interacting term among the $f$ fermions can be obtained by removing the bracket $\langle f | \cdots |f \rangle$. This result can be simply generalized to the  lattice version by extending the summation of $i$ and $j$ over the entire lattice, as shown in Eqs. (\ref{eq:HJ}) and (\ref{eq:J_H}).

\section{The mean-field Kondo-Heisenberg Hamiltonian on a strip} 
In this section, we provide the details of the matrix elements of the Kondo-Heisenberg Hamiltonian on a nano-strip with $N_y$ chains along $y$-axis. We choose the basis of the Kondo-Heisenberg strip as
\begin{widetext}
\begin{align}
    & \phi_{A,k}=\left(c_{k1\uparrow},c_{k2\uparrow},\cdots,c_{k N_y\uparrow}, c_{-k 1\uparrow}^{\dagger},c_{-k 2\uparrow}^{\dagger},\cdots,c_{-k N_y \uparrow}^{\dagger}, f_{k 1 \downarrow},f_{k 2 \downarrow},\cdots,f_{k N_y \downarrow}, f_{-k 1 \downarrow}^{\dagger},f_{-k 2\downarrow}^{\dagger},\cdots,f_{-k N_y \downarrow}^{\dagger}\right)^{T}, \nn
    &\phi_{B, k}=\left(c_{k 1\downarrow},c_{k 2\downarrow},\cdots,c_{k N_y\downarrow}, c_{-k 1\downarrow}^{\dagger},c_{-k 2\downarrow}^{\dagger},\cdots,c_{-k N_y \downarrow}^{\dagger}, f_{k 1 \uparrow},f_{k 2 \uparrow},\cdots,f_{k N_y \uparrow}, f_{-k 1 \uparrow}^{\dagger},f_{-k 2\uparrow}^{\dagger},\cdots,f_{-k N_y \uparrow}^{\dagger}\right)^{T},
\end{align}
\end{widetext}
where we take $k_{x}\to k$. The total Hamiltonian $H$ is represented as a summation of two decoupled Hamiltonians, $H_A$ and $H_B$, each of which is $4N_y \times 4N_y$ in size, given by
\begin{align}
    H = \sum_{k}\phi_{A,k}^{\dagger}\mathcal{H}_{A}(k)\phi_{A,k}+\sum_{k}\phi_{B,k}^{\dagger}\mathcal{H}_{B}(k)\phi_{B,k}.
\end{align}
Below, we provides the matrix elements of $\mathcal{H}_{A} (k)$ and $\mathcal{H}_{B}$, respectively: 
\subsection{$\mathcal{H}_A$}
The matrix elements of the hopping term for $\mathcal{H}_A$ are
\begin{align}
 &\mathcal{H}_{A}(y_i,y_i)=-t\cos k-\frac{\mu}{2}, \nn
 &\mathcal{H}_{A}(y_i+N_{y},y_i+N_{y})=t\cos k+\frac{\mu}{2}
\end{align}
for $y_i=1,2,\cdots,N_{y}$ while 
\begin{align}
	&\mathcal{H}_{A}(y_i,y_i+1)=-\frac{t}{2},\nn
	&\mathcal{H}_{A}(y_i+1,y_i)=-\frac{t}{2},\nn
    &\mathcal{H}_{A}(N_{y}+y_i+1,N_{y}+y_i)=\frac{t}{2},\nn
	& \mathcal{H}_{A}(N_{y}+y_i,N_{y}+y_i+1)=\frac{t}{2}
\end{align}
for $y_i=1,2,\cdots,N_{y}-1$.

For $H_{f}$, we have for $y_i=1,2,\cdots,N_{y}$ 
\begin{align}
    & \mathcal{H}_{A}(2N_{y}+y_i,2N_{y}+y_i)=\lambda/2, \nn
    & \mathcal{H}_{A}(3N_{y}+y_i,3N_{y}+y_i)=-\lambda/2.
\end{align}
The Kondo term $H_K$ for $\mathcal{H}_A$ describes the Kondo interaction with the following matrix form: the Kondo hybridization of $c$ and $f$ with the same $y$ chain are
\begin{align}
   &\mathcal{H}_{A}(2N_{y}+y_i,y_i)=x\sin k, \nn
   & \mathcal{H}_{A}(N_{y}+y_i,3N_{y}+y_i)=x\sin k,\nn
   & \mathcal{H}_{A}(y_i,y_i+2N_{y})=x\sin k, \nn
 & \mathcal{H}_{A}(3N_{y}+y_i,N_{y}+y_i)=x\sin k
\end{align}
for $y_i=1,\cdots,N_{y}$. The matrix elements of the Kondo term for $y_i=1,\cdots,N_{y}-1$ are
\begin{align}
    &\mathcal{H}_{A}(2N_{y}+y_i+1,y_i)=-\frac{x}{2},\nn
	&\mathcal{H}_{A}(N_{y}+y_i,3N_{y}+y_i+1)=\frac{x}{2}, \nn
	&\mathcal{H}_{A}(2N_{y}+y_i,y_i+1)=\frac{x}{2},\nn
	&\mathcal{H}_{A}(N_{y}+y_i+1,3N_{y}+y_i)=-\frac{x}{2},\nn
    &\mathcal{H}_{A}(y_i,2N_{y}+y_i+1)=-\frac{x}{2},\nn
	&\mathcal{H}_{A}(3N_{y}+y_i+1,N_{y}+y_i)=\frac{x}{2},\nn
	&\mathcal{H}_{A}(y_i+1,2N_{y}+y_i)=\frac{x}{2},\nn
	&\mathcal{H}_{A}(3N_{y}+y_i,N_{y}+y_i+1)=-\frac{x}{2},
\end{align}
which corresponds to the hybridization of $c$ and $f$ with the nearest-neighboring $y$ chains.

The RVB pairing term $H_{J}$ on a nano-strip is described by the following matrix elements: for $ y_i=1,\cdots,N_{y}$,
\begin{align}
    & \mathcal{H}_{\Delta}^{A}(2N_{y}+i,3N_{y}+i)=-i\Delta_{t}\sin k ,\nn
    & \mathcal{H}_{\Delta}^{A}(3N_{y}+i,2N_{y}+i)=i\Delta_{t}\sin k
\end{align}
are the matrix elements for the pairing of spinons with the same $y_i$. For $ y_i=1,\cdots,N_{y}-1$, we have
\begin{align}
    &\mathcal{H}_{A}(2N_{y}+y_i,3N_{y}+y_i+1)=-\frac{i}{2}\Delta_{t},\nn
    &\mathcal{H}_{A}(2N_{y}+y_i+1,3N_{y}+y_i)=\frac{i}{2}\Delta_{t}.\nn
    &\mathcal{H}_{A}(3N_{y}+y_i+1,2N_{y}+y_i)=\frac{i}{2}\Delta_{t},\nn
    &\mathcal{H}_{A}(3N_{y}+y_i,2N_{y}+y_i+1)=-\frac{i}{2}\Delta_{t}.
\end{align}

\subsection{$\mathcal{H}_B$}
 
The matrix elements for the hopping term in $H_B$ are
\begin{align}
    &\mathcal{H}_{B}(y_i,y_i)=-t\cos k-\frac{\mu}{2},\nn
    &\mathcal{H}_{B}(y_i+N_{y},y_i+N_{y})=t\cos k+\frac{\mu}{2}
\end{align}
for $y_i=1,2,\cdots,N_{y}$. While, for for $y_i=1,2,\cdots,N_{y}-1$, we obtain
\begin{align}
    &\mathcal{H}_{B}(y_i,y_i+1)=-\frac{t}{2},\nn 
    &\mathcal{H}_{B}(y_i+1,y_i)=-\frac{t}{2},\nn
    &\mathcal{H}_{B}(N_{y}+y_i+1,N_{y}+y_i)=\frac{t}{2},\nn 
    &\mathcal{H}_{B}(N_{y}+y_i,N_{y}+y_i+1)=\frac{t}{2}.
\end{align}
The matrix elements for $H_f$ are 
\begin{align}
  & \mathcal{H}_{B}(2N_{y}+y_i,2N_{y}+y_i)=\lambda/2,\nn 
  & \mathcal{H}_{B}(3N_{y}+y_i,3N_{y}+y_i)=-\lambda/2
\end{align}
with $y_i=1,2,\cdots,N_{y}$. 

The matrix elements of the Kondo term for $c$ and $f$ lying on the same chain $y_i$ are
\begin{align}
    &\mathcal{H}_{B}(2N_{y}+y_i,y_i)=x\sin k, \nn 
    &\mathcal{H}_{B}(N_{y}+y_i,3N_{y}+y_i)=x\sin k, \nn
    &\mathcal{H}_{B}(y_i,2N_{y}+y_i)=x\sin k, \nn 
    &\mathcal{H}_{B}(3N_{y}+y_i,N_{y}+y_i)=x\sin k,
\end{align}
where $y_i=1,\cdots,N_{y}$. For Kondo term where the hybridization is happening between nearest-neighboring chains, we have 
\begin{align}
    &\mathcal{H}_{B}(2N_{y}+y_i+1,y_i)=\frac{x}{2}, \nn
    &\mathcal{H}_{B}(N_{y}+y_i,3N_{y}+y_i+1)=-\frac{x}{2}\nn &\mathcal{H}_{B}(2N_{y}+y_i,y_i+1)=-\frac{x}{2},\nn &\mathcal{H}_{B}(N_{y}+y_i+1,3N_{y}+y_i)=\frac{x}{2}, \nn
    &\mathcal{H}_{B}(y_i,2N_{y}+y_i+1)=\frac{x}{2}\nn &\mathcal{H}_{B}(3N_{y}+y_i+1,N_{y}+y_i)=-\frac{x}{2},\nn &\mathcal{H}_{B}(y_i+1,2N_{y}+y_i)=-\frac{x}{2},\nn &\mathcal{H}_{B}(3N_{y}+y_i,N_{y}+y_i+1)=\frac{x}{2}
\end{align}
for $y_i=1,\cdots,N_{y}-1$. 

The matrix elements for the RVB spinon-pairing term are 
\begin{align}
   &  \mathcal{H}_{B}(2N_{y}+y_i,3N_{y}+y_i)=-i\Delta_{t}\sin k ,\nn
   & \mathcal{H}_{B}(3N_{y}+y_i,2N_{y}+y_i)=i\Delta_{t}\sin k 
\end{align}
for $y_i=1,\cdots,N_{y}$, and 
\begin{align}
    & \mathcal{H}_{B}(2N_{y}+y_i,3N_{y}+y_i+1)=\frac{i}{2}\Delta_{t},\nn
    & \mathcal{H}_{B}(2N_{y}+y_i+1,3N_{y}+y_i)=-\frac{i}{2}\Delta_{t}, \nn
    & \mathcal{H}_{B}(3N_{y}+y_i+1,2N_{y}+y_i)=-\frac{i}{2}\Delta_{t},\nn
    & \mathcal{H}_{B}(3N_{y}+y_i,2N_{y}+y_i+1)=\frac{i}{2}\Delta_{t}
\end{align}
for $y_i=1,\cdots,N_{y}-1$.


\begin{thebibliography}{54}%
\makeatletter
\providecommand \@ifxundefined [1]{%
 \@ifx{#1\undefined}
}%
\providecommand \@ifnum [1]{%
 \ifnum #1\expandafter \@firstoftwo
 \else \expandafter \@secondoftwo
 \fi
}%
\providecommand \@ifx [1]{%
 \ifx #1\expandafter \@firstoftwo
 \else \expandafter \@secondoftwo
 \fi
}%
\providecommand \natexlab [1]{#1}%
\providecommand \enquote  [1]{``#1''}%
\providecommand \bibnamefont  [1]{#1}%
\providecommand \bibfnamefont [1]{#1}%
\providecommand \citenamefont [1]{#1}%
\providecommand \href@noop [0]{\@secondoftwo}%
\providecommand \href [0]{\begingroup \@sanitize@url \@href}%
\providecommand \@href[1]{\@@startlink{#1}\@@href}%
\providecommand \@@href[1]{\endgroup#1\@@endlink}%
\providecommand \@sanitize@url [0]{\catcode `\\12\catcode `\$12\catcode
  `\&12\catcode `\#12\catcode `\^12\catcode `\_12\catcode `\%12\relax}%
\providecommand \@@startlink[1]{}%
\providecommand \@@endlink[0]{}%
\providecommand \url  [0]{\begingroup\@sanitize@url \@url }%
\providecommand \@url [1]{\endgroup\@href {#1}{\urlprefix }}%
\providecommand \urlprefix  [0]{URL }%
\providecommand \Eprint [0]{\href }%
\providecommand \doibase [0]{https://doi.org/}%
\providecommand \selectlanguage [0]{\@gobble}%
\providecommand \bibinfo  [0]{\@secondoftwo}%
\providecommand \bibfield  [0]{\@secondoftwo}%
\providecommand \translation [1]{[#1]}%
\providecommand \BibitemOpen [0]{}%
\providecommand \bibitemStop [0]{}%
\providecommand \bibitemNoStop [0]{.\EOS\space}%
\providecommand \EOS [0]{\spacefactor3000\relax}%
\providecommand \BibitemShut  [1]{\csname bibitem#1\endcsname}%
\let\auto@bib@innerbib\@empty
%</preamble>
\bibitem [{\citenamefont {Qi}\ and\ \citenamefont
  {Zhang}(2011)}]{Liang-RMP-2011}%
  \BibitemOpen
  \bibfield  {author} {\bibinfo {author} {\bibfnamefont {X.-L.}\ \bibnamefont
  {Qi}}\ and\ \bibinfo {author} {\bibfnamefont {S.-C.}\ \bibnamefont {Zhang}},\
  }\href {https://doi.org/10.1103/RevModPhys.83.1057} {\bibfield  {journal}
  {\bibinfo  {journal} {Rev. Mod. Phys.}\ }\textbf {\bibinfo {volume} {83}},\
  \bibinfo {pages} {1057} (\bibinfo {year} {2011})}\BibitemShut {NoStop}%
\bibitem [{\citenamefont {Alicea}(2012)}]{Alicea-MF-2012}%
  \BibitemOpen
  \bibfield  {author} {\bibinfo {author} {\bibfnamefont {J.}~\bibnamefont
  {Alicea}},\ }\href {https://doi.org/10.1088/0034-4885/75/7/076501} {\bibfield
   {journal} {\bibinfo  {journal} {Reports on Progress in Physics}\ }\textbf
  {\bibinfo {volume} {75}},\ \bibinfo {pages} {076501} (\bibinfo {year}
  {2012})}\BibitemShut {NoStop}%
\bibitem [{\citenamefont {Lutchyn}\ \emph {et~al.}(2010)\citenamefont
  {Lutchyn}, \citenamefont {Sau},\ and\ \citenamefont
  {Das~Sarma}}]{Sau-PRL-MF}%
  \BibitemOpen
  \bibfield  {author} {\bibinfo {author} {\bibfnamefont {R.~M.}\ \bibnamefont
  {Lutchyn}}, \bibinfo {author} {\bibfnamefont {J.~D.}\ \bibnamefont {Sau}},\
  and\ \bibinfo {author} {\bibfnamefont {S.}~\bibnamefont {Das~Sarma}},\ }\href
  {https://doi.org/10.1103/PhysRevLett.105.077001} {\bibfield  {journal}
  {\bibinfo  {journal} {Phys. Rev. Lett.}\ }\textbf {\bibinfo {volume} {105}},\
  \bibinfo {pages} {077001} (\bibinfo {year} {2010})}\BibitemShut {NoStop}%
\bibitem [{\citenamefont {Oreg}\ \emph {et~al.}(2010)\citenamefont {Oreg},
  \citenamefont {Refael},\ and\ \citenamefont {von Oppen}}]{Oreg-PRL-MF}%
  \BibitemOpen
  \bibfield  {author} {\bibinfo {author} {\bibfnamefont {Y.}~\bibnamefont
  {Oreg}}, \bibinfo {author} {\bibfnamefont {G.}~\bibnamefont {Refael}},\ and\
  \bibinfo {author} {\bibfnamefont {F.}~\bibnamefont {von Oppen}},\ }\href
  {https://doi.org/10.1103/PhysRevLett.105.177002} {\bibfield  {journal}
  {\bibinfo  {journal} {Phys. Rev. Lett.}\ }\textbf {\bibinfo {volume} {105}},\
  \bibinfo {pages} {177002} (\bibinfo {year} {2010})}\BibitemShut {NoStop}%
\bibitem [{\citenamefont {Gaidamauskas}\ \emph {et~al.}(2014)\citenamefont
  {Gaidamauskas}, \citenamefont {Paaske},\ and\ \citenamefont
  {Flensberg}}]{Paaske-PRL-MF}%
  \BibitemOpen
  \bibfield  {author} {\bibinfo {author} {\bibfnamefont {E.}~\bibnamefont
  {Gaidamauskas}}, \bibinfo {author} {\bibfnamefont {J.}~\bibnamefont
  {Paaske}},\ and\ \bibinfo {author} {\bibfnamefont {K.}~\bibnamefont
  {Flensberg}},\ }\href {https://doi.org/10.1103/PhysRevLett.112.126402}
  {\bibfield  {journal} {\bibinfo  {journal} {Phys. Rev. Lett.}\ }\textbf
  {\bibinfo {volume} {112}},\ \bibinfo {pages} {126402} (\bibinfo {year}
  {2014})}\BibitemShut {NoStop}%
\bibitem [{\citenamefont {Dzero}\ \emph {et~al.}(2016)\citenamefont {Dzero},
  \citenamefont {Xia}, \citenamefont {Galitski},\ and\ \citenamefont
  {Coleman}}]{dzero-Ann-TKI}%
  \BibitemOpen
  \bibfield  {author} {\bibinfo {author} {\bibfnamefont {M.}~\bibnamefont
  {Dzero}}, \bibinfo {author} {\bibfnamefont {J.}~\bibnamefont {Xia}}, \bibinfo
  {author} {\bibfnamefont {V.}~\bibnamefont {Galitski}},\ and\ \bibinfo
  {author} {\bibfnamefont {P.}~\bibnamefont {Coleman}},\ }\href
  {https://doi.org/10.1146/annurev-conmatphys-031214-014749} {\bibfield
  {journal} {\bibinfo  {journal} {Annual Review of Condensed Matter Physics}\
  }\textbf {\bibinfo {volume} {7}},\ \bibinfo {pages} {249} (\bibinfo {year}
  {2016})}\BibitemShut {NoStop}%
\bibitem [{\citenamefont {Dzero}\ \emph {et~al.}(2012)\citenamefont {Dzero},
  \citenamefont {Sun}, \citenamefont {Coleman},\ and\ \citenamefont
  {Galitski}}]{dzero-TKI-PRB}%
  \BibitemOpen
  \bibfield  {author} {\bibinfo {author} {\bibfnamefont {M.}~\bibnamefont
  {Dzero}}, \bibinfo {author} {\bibfnamefont {K.}~\bibnamefont {Sun}}, \bibinfo
  {author} {\bibfnamefont {P.}~\bibnamefont {Coleman}},\ and\ \bibinfo {author}
  {\bibfnamefont {V.}~\bibnamefont {Galitski}},\ }\href
  {https://doi.org/10.1103/PhysRevB.85.045130} {\bibfield  {journal} {\bibinfo
  {journal} {Phys. Rev. B}\ }\textbf {\bibinfo {volume} {85}},\ \bibinfo
  {pages} {045130} (\bibinfo {year} {2012})}\BibitemShut {NoStop}%
\bibitem [{\citenamefont {Dzero}\ \emph {et~al.}(2010)\citenamefont {Dzero},
  \citenamefont {Sun}, \citenamefont {Galitski},\ and\ \citenamefont
  {Coleman}}]{dzero-TKI-PRL}%
  \BibitemOpen
  \bibfield  {author} {\bibinfo {author} {\bibfnamefont {M.}~\bibnamefont
  {Dzero}}, \bibinfo {author} {\bibfnamefont {K.}~\bibnamefont {Sun}}, \bibinfo
  {author} {\bibfnamefont {V.}~\bibnamefont {Galitski}},\ and\ \bibinfo
  {author} {\bibfnamefont {P.}~\bibnamefont {Coleman}},\ }\href
  {https://doi.org/10.1103/PhysRevLett.104.106408} {\bibfield  {journal}
  {\bibinfo  {journal} {Phys. Rev. Lett.}\ }\textbf {\bibinfo {volume} {104}},\
  \bibinfo {pages} {106408} (\bibinfo {year} {2010})}\BibitemShut {NoStop}%
\bibitem [{\citenamefont {Lai}\ \emph {et~al.}(2018)\citenamefont {Lai},
  \citenamefont {Grefe}, \citenamefont {Paschen},\ and\ \citenamefont
  {Si}}]{Weyl-kondo-PNAS}%
  \BibitemOpen
  \bibfield  {author} {\bibinfo {author} {\bibfnamefont {H.-H.}\ \bibnamefont
  {Lai}}, \bibinfo {author} {\bibfnamefont {S.~E.}\ \bibnamefont {Grefe}},
  \bibinfo {author} {\bibfnamefont {S.}~\bibnamefont {Paschen}},\ and\ \bibinfo
  {author} {\bibfnamefont {Q.}~\bibnamefont {Si}},\ }\href
  {https://doi.org/10.1073/pnas.1715851115} {\bibfield  {journal} {\bibinfo
  {journal} {Proceedings of the National Academy of Sciences}\ }\textbf
  {\bibinfo {volume} {115}},\ \bibinfo {pages} {93} (\bibinfo {year}
  {2018})}\BibitemShut {NoStop}%
\bibitem [{\citenamefont {Mackenzie}\ and\ \citenamefont
  {Maeno}(2003)}]{Mackenzie-RMP-SrRuO}%
  \BibitemOpen
  \bibfield  {author} {\bibinfo {author} {\bibfnamefont {A.~P.}\ \bibnamefont
  {Mackenzie}}\ and\ \bibinfo {author} {\bibfnamefont {Y.}~\bibnamefont
  {Maeno}},\ }\href {https://doi.org/10.1103/RevModPhys.75.657} {\bibfield
  {journal} {\bibinfo  {journal} {Rev. Mod. Phys.}\ }\textbf {\bibinfo {volume}
  {75}},\ \bibinfo {pages} {657} (\bibinfo {year} {2003})}\BibitemShut
  {NoStop}%
\bibitem [{\citenamefont {Maeno}\ \emph {et~al.}(2012)\citenamefont {Maeno},
  \citenamefont {Kittaka}, \citenamefont {Nomura}, \citenamefont {Yonezawa},\
  and\ \citenamefont {Ishida}}]{Maeno-JPSJ-SrRuO}%
  \BibitemOpen
  \bibfield  {author} {\bibinfo {author} {\bibfnamefont {Y.}~\bibnamefont
  {Maeno}}, \bibinfo {author} {\bibfnamefont {S.}~\bibnamefont {Kittaka}},
  \bibinfo {author} {\bibfnamefont {T.}~\bibnamefont {Nomura}}, \bibinfo
  {author} {\bibfnamefont {S.}~\bibnamefont {Yonezawa}},\ and\ \bibinfo
  {author} {\bibfnamefont {K.}~\bibnamefont {Ishida}},\ }\href
  {https://doi.org/10.1143/JPSJ.81.011009} {\bibfield  {journal} {\bibinfo
  {journal} {Journal of the Physical Society of Japan}\ }\textbf {\bibinfo
  {volume} {81}},\ \bibinfo {pages} {011009} (\bibinfo {year} {2012})},\
  \Eprint {https://arxiv.org/abs/https://doi.org/10.1143/JPSJ.81.011009}
  {https://doi.org/10.1143/JPSJ.81.011009} \BibitemShut {NoStop}%
\bibitem [{\citenamefont {Kallin}\ and\ \citenamefont
  {Berlinsky}(2009)}]{Kallin-JPCM-2009}%
  \BibitemOpen
  \bibfield  {author} {\bibinfo {author} {\bibfnamefont {C.}~\bibnamefont
  {Kallin}}\ and\ \bibinfo {author} {\bibfnamefont {A.~J.}\ \bibnamefont
  {Berlinsky}},\ }\href {https://doi.org/10.1088/0953-8984/21/16/164210}
  {\bibfield  {journal} {\bibinfo  {journal} {Journal of Physics: Condensed
  Matter}\ }\textbf {\bibinfo {volume} {21}},\ \bibinfo {pages} {164210}
  (\bibinfo {year} {2009})}\BibitemShut {NoStop}%
\bibitem [{\citenamefont {Xu}\ \emph {et~al.}(2020)\citenamefont {Xu},
  \citenamefont {Li},\ and\ \citenamefont {Chien}}]{CLChen-BiPd-PRL}%
  \BibitemOpen
  \bibfield  {author} {\bibinfo {author} {\bibfnamefont {X.}~\bibnamefont
  {Xu}}, \bibinfo {author} {\bibfnamefont {Y.}~\bibnamefont {Li}},\ and\
  \bibinfo {author} {\bibfnamefont {C.~L.}\ \bibnamefont {Chien}},\ }\href
  {https://doi.org/10.1103/PhysRevLett.124.167001} {\bibfield  {journal}
  {\bibinfo  {journal} {Phys. Rev. Lett.}\ }\textbf {\bibinfo {volume} {124}},\
  \bibinfo {pages} {167001} (\bibinfo {year} {2020})}\BibitemShut {NoStop}%
\bibitem [{\citenamefont {Ran}\ \emph {et~al.}(2019{\natexlab{a}})\citenamefont
  {Ran}, \citenamefont {Eckberg}, \citenamefont {Ding}, \citenamefont
  {Furukawa}, \citenamefont {Metz}, \citenamefont {Saha}, \citenamefont {Liu},
  \citenamefont {Zic}, \citenamefont {Kim}, \citenamefont {Paglione},\ and\
  \citenamefont {Butch}}]{Ran-FM-UTe}%
  \BibitemOpen
  \bibfield  {author} {\bibinfo {author} {\bibfnamefont {S.}~\bibnamefont
  {Ran}}, \bibinfo {author} {\bibfnamefont {C.}~\bibnamefont {Eckberg}},
  \bibinfo {author} {\bibfnamefont {Q.-P.}\ \bibnamefont {Ding}}, \bibinfo
  {author} {\bibfnamefont {Y.}~\bibnamefont {Furukawa}}, \bibinfo {author}
  {\bibfnamefont {T.}~\bibnamefont {Metz}}, \bibinfo {author} {\bibfnamefont
  {S.~R.}\ \bibnamefont {Saha}}, \bibinfo {author} {\bibfnamefont {I.-L.}\
  \bibnamefont {Liu}}, \bibinfo {author} {\bibfnamefont {M.}~\bibnamefont
  {Zic}}, \bibinfo {author} {\bibfnamefont {H.}~\bibnamefont {Kim}}, \bibinfo
  {author} {\bibfnamefont {J.}~\bibnamefont {Paglione}},\ and\ \bibinfo
  {author} {\bibfnamefont {N.~P.}\ \bibnamefont {Butch}},\ }\href
  {https://doi.org/10.1126/science.aav8645} {\bibfield  {journal} {\bibinfo
  {journal} {Science}\ }\textbf {\bibinfo {volume} {365}},\ \bibinfo {pages}
  {684} (\bibinfo {year} {2019}{\natexlab{a}})},\ \Eprint
  {https://arxiv.org/abs/https://www.science.org/doi/pdf/10.1126/science.aav8645}
  {https://www.science.org/doi/pdf/10.1126/science.aav8645} \BibitemShut
  {NoStop}%
\bibitem [{\citenamefont {Ran}\ \emph {et~al.}(2019{\natexlab{b}})\citenamefont
  {Ran}, \citenamefont {Liu}, \citenamefont {Eo}, \citenamefont {Campbell},
  \citenamefont {Neves}, \citenamefont {Fuhrman}, \citenamefont {Saha},
  \citenamefont {Eckberg}, \citenamefont {Kim}, \citenamefont {Graf},
  \citenamefont {Balakirev}, \citenamefont {Singleton}, \citenamefont
  {Paglione},\ and\ \citenamefont {Butch}}]{Ran-UTe-Nature-ExtremeMag}%
  \BibitemOpen
  \bibfield  {author} {\bibinfo {author} {\bibfnamefont {S.}~\bibnamefont
  {Ran}}, \bibinfo {author} {\bibfnamefont {I.-L.}\ \bibnamefont {Liu}},
  \bibinfo {author} {\bibfnamefont {Y.~S.}\ \bibnamefont {Eo}}, \bibinfo
  {author} {\bibfnamefont {D.~J.}\ \bibnamefont {Campbell}}, \bibinfo {author}
  {\bibfnamefont {P.~M.}\ \bibnamefont {Neves}}, \bibinfo {author}
  {\bibfnamefont {W.~T.}\ \bibnamefont {Fuhrman}}, \bibinfo {author}
  {\bibfnamefont {S.~R.}\ \bibnamefont {Saha}}, \bibinfo {author}
  {\bibfnamefont {C.}~\bibnamefont {Eckberg}}, \bibinfo {author} {\bibfnamefont
  {H.}~\bibnamefont {Kim}}, \bibinfo {author} {\bibfnamefont {D.}~\bibnamefont
  {Graf}}, \bibinfo {author} {\bibfnamefont {F.}~\bibnamefont {Balakirev}},
  \bibinfo {author} {\bibfnamefont {J.}~\bibnamefont {Singleton}}, \bibinfo
  {author} {\bibfnamefont {J.}~\bibnamefont {Paglione}},\ and\ \bibinfo
  {author} {\bibfnamefont {N.~P.}\ \bibnamefont {Butch}},\ }\href
  {https://doi.org/10.1038/s41567-019-0670-x} {\bibfield  {journal} {\bibinfo
  {journal} {Nature Physics}\ }\textbf {\bibinfo {volume} {15}},\ \bibinfo
  {pages} {1250} (\bibinfo {year} {2019}{\natexlab{b}})}\BibitemShut {NoStop}%
\bibitem [{\citenamefont {Aoki}\ \emph {et~al.}(2019)\citenamefont {Aoki},
  \citenamefont {Nakamura}, \citenamefont {Honda}, \citenamefont {Li},
  \citenamefont {Homma}, \citenamefont {Shimizu}, \citenamefont {Sato},
  \citenamefont {Knebel}, \citenamefont {Brison}, \citenamefont {Pourret},
  \citenamefont {Braithwaite}, \citenamefont {Lapertot}, \citenamefont {Niu},
  \citenamefont {Vali\v{s}ka}, \citenamefont {Harima},\ and\ \citenamefont
  {Flouquet}}]{Aoki-2019-JPSJ}%
  \BibitemOpen
  \bibfield  {author} {\bibinfo {author} {\bibfnamefont {D.}~\bibnamefont
  {Aoki}}, \bibinfo {author} {\bibfnamefont {A.}~\bibnamefont {Nakamura}},
  \bibinfo {author} {\bibfnamefont {F.}~\bibnamefont {Honda}}, \bibinfo
  {author} {\bibfnamefont {D.}~\bibnamefont {Li}}, \bibinfo {author}
  {\bibfnamefont {Y.}~\bibnamefont {Homma}}, \bibinfo {author} {\bibfnamefont
  {Y.}~\bibnamefont {Shimizu}}, \bibinfo {author} {\bibfnamefont {Y.~J.}\
  \bibnamefont {Sato}}, \bibinfo {author} {\bibfnamefont {G.}~\bibnamefont
  {Knebel}}, \bibinfo {author} {\bibfnamefont {J.-P.}\ \bibnamefont {Brison}},
  \bibinfo {author} {\bibfnamefont {A.}~\bibnamefont {Pourret}}, \bibinfo
  {author} {\bibfnamefont {D.}~\bibnamefont {Braithwaite}}, \bibinfo {author}
  {\bibfnamefont {G.}~\bibnamefont {Lapertot}}, \bibinfo {author}
  {\bibfnamefont {Q.}~\bibnamefont {Niu}}, \bibinfo {author} {\bibfnamefont
  {M.}~\bibnamefont {Vali\v{s}ka}}, \bibinfo {author} {\bibfnamefont
  {H.}~\bibnamefont {Harima}},\ and\ \bibinfo {author} {\bibfnamefont
  {J.}~\bibnamefont {Flouquet}},\ }\href
  {https://doi.org/10.7566/JPSJ.88.043702} {\bibfield  {journal} {\bibinfo
  {journal} {Journal of the Physical Society of Japan}\ }\textbf {\bibinfo
  {volume} {88}},\ \bibinfo {pages} {043702} (\bibinfo {year}
  {2019})}\BibitemShut {NoStop}%
\bibitem [{\citenamefont {Jiao}\ \emph {et~al.}(2020)\citenamefont {Jiao},
  \citenamefont {Howard}, \citenamefont {Ran}, \citenamefont {Wang},
  \citenamefont {Rodriguez}, \citenamefont {Sigrist}, \citenamefont {Wang},
  \citenamefont {Butch},\ and\ \citenamefont {Madhavan}}]{Jiao-2020-UTe2}%
  \BibitemOpen
  \bibfield  {author} {\bibinfo {author} {\bibfnamefont {L.}~\bibnamefont
  {Jiao}}, \bibinfo {author} {\bibfnamefont {S.}~\bibnamefont {Howard}},
  \bibinfo {author} {\bibfnamefont {S.}~\bibnamefont {Ran}}, \bibinfo {author}
  {\bibfnamefont {Z.}~\bibnamefont {Wang}}, \bibinfo {author} {\bibfnamefont
  {J.~O.}\ \bibnamefont {Rodriguez}}, \bibinfo {author} {\bibfnamefont
  {M.}~\bibnamefont {Sigrist}}, \bibinfo {author} {\bibfnamefont
  {Z.}~\bibnamefont {Wang}}, \bibinfo {author} {\bibfnamefont {N.~P.}\
  \bibnamefont {Butch}},\ and\ \bibinfo {author} {\bibfnamefont
  {V.}~\bibnamefont {Madhavan}},\ }\href
  {https://doi.org/10.1038/s41586-020-2122-2} {\bibfield  {journal} {\bibinfo
  {journal} {Nature}\ }\textbf {\bibinfo {volume} {579}},\ \bibinfo {pages}
  {523} (\bibinfo {year} {2020})}\BibitemShut {NoStop}%
\bibitem [{\citenamefont {Choi}\ \emph {et~al.}(2018)\citenamefont {Choi},
  \citenamefont {Klein}, \citenamefont {Rosch},\ and\ \citenamefont
  {Kim}}]{Kim-Kondo-Kitaev}%
  \BibitemOpen
  \bibfield  {author} {\bibinfo {author} {\bibfnamefont {W.}~\bibnamefont
  {Choi}}, \bibinfo {author} {\bibfnamefont {P.~W.}\ \bibnamefont {Klein}},
  \bibinfo {author} {\bibfnamefont {A.}~\bibnamefont {Rosch}},\ and\ \bibinfo
  {author} {\bibfnamefont {Y.~B.}\ \bibnamefont {Kim}},\ }\href
  {https://doi.org/10.1103/PhysRevB.98.155123} {\bibfield  {journal} {\bibinfo
  {journal} {Phys. Rev. B}\ }\textbf {\bibinfo {volume} {98}},\ \bibinfo
  {pages} {155123} (\bibinfo {year} {2018})}\BibitemShut {NoStop}%
\bibitem [{\citenamefont {Schrieffer}\ and\ \citenamefont
  {Wolff}(1966)}]{SW-transformation}%
  \BibitemOpen
  \bibfield  {author} {\bibinfo {author} {\bibfnamefont {J.~R.}\ \bibnamefont
  {Schrieffer}}\ and\ \bibinfo {author} {\bibfnamefont {P.~A.}\ \bibnamefont
  {Wolff}},\ }\href {https://doi.org/10.1103/PhysRev.149.491} {\bibfield
  {journal} {\bibinfo  {journal} {Phys. Rev.}\ }\textbf {\bibinfo {volume}
  {149}},\ \bibinfo {pages} {491} (\bibinfo {year} {1966})}\BibitemShut
  {NoStop}%
\bibitem [{\citenamefont {Hewson}(1997)}]{hewson1997kondo}%
  \BibitemOpen
  \bibfield  {author} {\bibinfo {author} {\bibfnamefont {A.~C.}\ \bibnamefont
  {Hewson}},\ }\href@noop {} {\emph {\bibinfo {title} {The Kondo problem to
  heavy fermions}}},\ Vol.~\bibinfo {volume} {2}\ (\bibinfo  {publisher}
  {Cambridge university press},\ \bibinfo {year} {1997})\BibitemShut {NoStop}%
\bibitem [{\citenamefont {Doniach}(1977)}]{Doniach-scenario}%
  \BibitemOpen
  \bibfield  {author} {\bibinfo {author} {\bibfnamefont {S.}~\bibnamefont
  {Doniach}},\ }\href
  {https://doi.org/https://doi.org/10.1016/0378-4363(77)90190-5} {\bibfield
  {journal} {\bibinfo  {journal} {Physica B+C}\ }\textbf {\bibinfo {volume}
  {91}},\ \bibinfo {pages} {231} (\bibinfo {year} {1977})}\BibitemShut
  {NoStop}%
\bibitem [{\citenamefont {Legner}(2016)}]{ETH-Phd-thesis}%
  \BibitemOpen
  \bibfield  {author} {\bibinfo {author} {\bibfnamefont {M.}~\bibnamefont
  {Legner}},\ }\emph {\bibinfo {title} {{Topological Kondo insulators:
  materials at the interface of topology and strong correlations}}},\ \href
  {https://doi.org/10.3929/ethz-a-010779690} {\bibinfo {type} {Doctoral
  thesis}},\ \bibinfo  {school} {ETH Zurich}, \bibinfo {address} {Zürich}
  (\bibinfo {year} {2016})\BibitemShut {NoStop}%
\bibitem [{\citenamefont {Ruderman}\ and\ \citenamefont
  {Kittel}(1954)}]{rkky-kittel}%
  \BibitemOpen
  \bibfield  {author} {\bibinfo {author} {\bibfnamefont {M.~A.}\ \bibnamefont
  {Ruderman}}\ and\ \bibinfo {author} {\bibfnamefont {C.}~\bibnamefont
  {Kittel}},\ }\href {https://doi.org/10.1103/PhysRev.96.99} {\bibfield
  {journal} {\bibinfo  {journal} {Phys. Rev.}\ }\textbf {\bibinfo {volume}
  {96}},\ \bibinfo {pages} {99} (\bibinfo {year} {1954})}\BibitemShut {NoStop}%
\bibitem [{\citenamefont {Van~Vleck}(1962)}]{rkky-vleck}%
  \BibitemOpen
  \bibfield  {author} {\bibinfo {author} {\bibfnamefont {J.~H.}\ \bibnamefont
  {Van~Vleck}},\ }\href {https://doi.org/10.1103/RevModPhys.34.681} {\bibfield
  {journal} {\bibinfo  {journal} {Rev. Mod. Phys.}\ }\textbf {\bibinfo {volume}
  {34}},\ \bibinfo {pages} {681} (\bibinfo {year} {1962})}\BibitemShut
  {NoStop}%
\bibitem [{\citenamefont {Kirchner}\ \emph {et~al.}(2020)\citenamefont
  {Kirchner}, \citenamefont {Paschen}, \citenamefont {Chen}, \citenamefont
  {Wirth}, \citenamefont {Feng}, \citenamefont {Thompson},\ and\ \citenamefont
  {Si}}]{RMP-Kirchner}%
  \BibitemOpen
  \bibfield  {author} {\bibinfo {author} {\bibfnamefont {S.}~\bibnamefont
  {Kirchner}}, \bibinfo {author} {\bibfnamefont {S.}~\bibnamefont {Paschen}},
  \bibinfo {author} {\bibfnamefont {Q.}~\bibnamefont {Chen}}, \bibinfo {author}
  {\bibfnamefont {S.}~\bibnamefont {Wirth}}, \bibinfo {author} {\bibfnamefont
  {D.}~\bibnamefont {Feng}}, \bibinfo {author} {\bibfnamefont {J.~D.}\
  \bibnamefont {Thompson}},\ and\ \bibinfo {author} {\bibfnamefont
  {Q.}~\bibnamefont {Si}},\ }\href
  {https://doi.org/10.1103/RevModPhys.92.011002} {\bibfield  {journal}
  {\bibinfo  {journal} {Rev. Mod. Phys.}\ }\textbf {\bibinfo {volume} {92}},\
  \bibinfo {pages} {011002} (\bibinfo {year} {2020})}\BibitemShut {NoStop}%
\bibitem [{\citenamefont {Wang}\ \emph {et~al.}(2022)\citenamefont {Wang},
  \citenamefont {Chang},\ and\ \citenamefont {Chung}}]{sm-phase-PNAS}%
  \BibitemOpen
  \bibfield  {author} {\bibinfo {author} {\bibfnamefont {J.}~\bibnamefont
  {Wang}}, \bibinfo {author} {\bibfnamefont {Y.-Y.}\ \bibnamefont {Chang}},\
  and\ \bibinfo {author} {\bibfnamefont {C.-H.}\ \bibnamefont {Chung}},\ }\href
  {https://doi.org/10.1073/pnas.2116980119} {\bibfield  {journal} {\bibinfo
  {journal} {Proceedings of the National Academy of Sciences}\ }\textbf
  {\bibinfo {volume} {119}},\ \bibinfo {pages} {e2116980119} (\bibinfo {year}
  {2022})}\BibitemShut {NoStop}%
\bibitem [{\citenamefont {Mineev}\ \emph {et~al.}(1999)\citenamefont {Mineev},
  \citenamefont {Samokhin}, \citenamefont {Landau},\ and\ \citenamefont
  {Landau}}]{Mineev-SC}%
  \BibitemOpen
  \bibfield  {author} {\bibinfo {author} {\bibfnamefont {V.}~\bibnamefont
  {Mineev}}, \bibinfo {author} {\bibfnamefont {K.}~\bibnamefont {Samokhin}},
  \bibinfo {author} {\bibfnamefont {L.}~\bibnamefont {Landau}},\ and\ \bibinfo
  {author} {\bibfnamefont {L.}~\bibnamefont {Landau}},\ }\href
  {https://books.google.com.tw/books?id=2BXYWT8m068C} {\emph {\bibinfo {title}
  {Introduction to Unconventional Superconductivity}}}\ (\bibinfo  {publisher}
  {Taylor \& Francis},\ \bibinfo {year} {1999})\BibitemShut {NoStop}%
\bibitem [{\citenamefont {Coleman}(2015)}]{coleman-book}%
  \BibitemOpen
  \bibfield  {author} {\bibinfo {author} {\bibfnamefont {P.}~\bibnamefont
  {Coleman}},\ }\href {https://doi.org/10.1017/CBO9781139020916} {\emph
  {\bibinfo {title} {Introduction to Many-Body Physics}}}\ (\bibinfo
  {publisher} {Cambridge University Press},\ \bibinfo {year}
  {2015})\BibitemShut {NoStop}%
\bibitem [{\citenamefont {Schnyder}\ \emph {et~al.}(2008)\citenamefont
  {Schnyder}, \citenamefont {Ryu}, \citenamefont {Furusaki},\ and\
  \citenamefont {Ludwig}}]{Shinsei-PRB-classification}%
  \BibitemOpen
  \bibfield  {author} {\bibinfo {author} {\bibfnamefont {A.~P.}\ \bibnamefont
  {Schnyder}}, \bibinfo {author} {\bibfnamefont {S.}~\bibnamefont {Ryu}},
  \bibinfo {author} {\bibfnamefont {A.}~\bibnamefont {Furusaki}},\ and\
  \bibinfo {author} {\bibfnamefont {A.~W.~W.}\ \bibnamefont {Ludwig}},\ }\href
  {https://doi.org/10.1103/PhysRevB.78.195125} {\bibfield  {journal} {\bibinfo
  {journal} {Phys. Rev. B}\ }\textbf {\bibinfo {volume} {78}},\ \bibinfo
  {pages} {195125} (\bibinfo {year} {2008})}\BibitemShut {NoStop}%
\bibitem [{\citenamefont {Coleman}\ and\ \citenamefont
  {Andrei}(1989)}]{Coleman-Andrei}%
  \BibitemOpen
  \bibfield  {author} {\bibinfo {author} {\bibfnamefont {P.}~\bibnamefont
  {Coleman}}\ and\ \bibinfo {author} {\bibfnamefont {N.}~\bibnamefont
  {Andrei}},\ }\href {https://doi.org/10.1088/0953-8984/1/26/003} {\bibfield
  {journal} {\bibinfo  {journal} {Journal of Physics: Condensed Matter}\
  }\textbf {\bibinfo {volume} {1}},\ \bibinfo {pages} {4057} (\bibinfo {year}
  {1989})}\BibitemShut {NoStop}%
\bibitem [{\citenamefont {Coleman}\ and\ \citenamefont
  {Nevidomskyy}(2010)}]{Nevidomskyy}%
  \BibitemOpen
  \bibfield  {author} {\bibinfo {author} {\bibfnamefont {P.}~\bibnamefont
  {Coleman}}\ and\ \bibinfo {author} {\bibfnamefont {A.~H.}\ \bibnamefont
  {Nevidomskyy}},\ }\href@noop {} {\bibfield  {journal} {\bibinfo  {journal}
  {Journal of Low Temperature Physics}\ }\textbf {\bibinfo {volume} {161}},\
  \bibinfo {pages} {182} (\bibinfo {year} {2010})}\BibitemShut {NoStop}%
\bibitem [{\citenamefont {Senthil}\ \emph {et~al.}(2003)\citenamefont
  {Senthil}, \citenamefont {Sachdev},\ and\ \citenamefont
  {Vojta}}]{Senthil-PRL-2003-fractionalized}%
  \BibitemOpen
  \bibfield  {author} {\bibinfo {author} {\bibfnamefont {T.}~\bibnamefont
  {Senthil}}, \bibinfo {author} {\bibfnamefont {S.}~\bibnamefont {Sachdev}},\
  and\ \bibinfo {author} {\bibfnamefont {M.}~\bibnamefont {Vojta}},\ }\href
  {https://doi.org/10.1103/PhysRevLett.90.216403} {\bibfield  {journal}
  {\bibinfo  {journal} {Phys. Rev. Lett.}\ }\textbf {\bibinfo {volume} {90}},\
  \bibinfo {pages} {216403} (\bibinfo {year} {2003})}\BibitemShut {NoStop}%
\bibitem [{\citenamefont {Fu}\ and\ \citenamefont
  {Kane}(2006)}]{Fu-kane-PRB-2005}%
  \BibitemOpen
  \bibfield  {author} {\bibinfo {author} {\bibfnamefont {L.}~\bibnamefont
  {Fu}}\ and\ \bibinfo {author} {\bibfnamefont {C.~L.}\ \bibnamefont {Kane}},\
  }\href {https://doi.org/10.1103/PhysRevB.74.195312} {\bibfield  {journal}
  {\bibinfo  {journal} {Phys. Rev. B}\ }\textbf {\bibinfo {volume} {74}},\
  \bibinfo {pages} {195312} (\bibinfo {year} {2006})}\BibitemShut {NoStop}%
\bibitem [{\citenamefont {Fu}\ \emph {et~al.}(2007)\citenamefont {Fu},
  \citenamefont {Kane},\ and\ \citenamefont {Mele}}]{Fu-kane-PRL-2007}%
  \BibitemOpen
  \bibfield  {author} {\bibinfo {author} {\bibfnamefont {L.}~\bibnamefont
  {Fu}}, \bibinfo {author} {\bibfnamefont {C.~L.}\ \bibnamefont {Kane}},\ and\
  \bibinfo {author} {\bibfnamefont {E.~J.}\ \bibnamefont {Mele}},\ }\href
  {https://doi.org/10.1103/PhysRevLett.98.106803} {\bibfield  {journal}
  {\bibinfo  {journal} {Phys. Rev. Lett.}\ }\textbf {\bibinfo {volume} {98}},\
  \bibinfo {pages} {106803} (\bibinfo {year} {2007})}\BibitemShut {NoStop}%
\bibitem [{\citenamefont {Sheng}\ \emph {et~al.}(2006)\citenamefont {Sheng},
  \citenamefont {Weng}, \citenamefont {Sheng},\ and\ \citenamefont
  {Haldane}}]{DNSheng-PRL-2006}%
  \BibitemOpen
  \bibfield  {author} {\bibinfo {author} {\bibfnamefont {D.~N.}\ \bibnamefont
  {Sheng}}, \bibinfo {author} {\bibfnamefont {Z.~Y.}\ \bibnamefont {Weng}},
  \bibinfo {author} {\bibfnamefont {L.}~\bibnamefont {Sheng}},\ and\ \bibinfo
  {author} {\bibfnamefont {F.~D.~M.}\ \bibnamefont {Haldane}},\ }\href
  {https://doi.org/10.1103/PhysRevLett.97.036808} {\bibfield  {journal}
  {\bibinfo  {journal} {Phys. Rev. Lett.}\ }\textbf {\bibinfo {volume} {97}},\
  \bibinfo {pages} {036808} (\bibinfo {year} {2006})}\BibitemShut {NoStop}%
\bibitem [{\citenamefont {Thouless}\ \emph {et~al.}(1982)\citenamefont
  {Thouless}, \citenamefont {Kohmoto}, \citenamefont {Nightingale},\ and\
  \citenamefont {den Nijs}}]{TKNN-PRL}%
  \BibitemOpen
  \bibfield  {author} {\bibinfo {author} {\bibfnamefont {D.~J.}\ \bibnamefont
  {Thouless}}, \bibinfo {author} {\bibfnamefont {M.}~\bibnamefont {Kohmoto}},
  \bibinfo {author} {\bibfnamefont {M.~P.}\ \bibnamefont {Nightingale}},\ and\
  \bibinfo {author} {\bibfnamefont {M.}~\bibnamefont {den Nijs}},\ }\href
  {https://doi.org/10.1103/PhysRevLett.49.405} {\bibfield  {journal} {\bibinfo
  {journal} {Phys. Rev. Lett.}\ }\textbf {\bibinfo {volume} {49}},\ \bibinfo
  {pages} {405} (\bibinfo {year} {1982})}\BibitemShut {NoStop}%
\bibitem [{\citenamefont {Fukui}\ \emph {et~al.}(2005)\citenamefont {Fukui},
  \citenamefont {Hatsugai},\ and\ \citenamefont
  {Suzuki}}]{Hatsugui-chern-numb}%
  \BibitemOpen
  \bibfield  {author} {\bibinfo {author} {\bibfnamefont {T.}~\bibnamefont
  {Fukui}}, \bibinfo {author} {\bibfnamefont {Y.}~\bibnamefont {Hatsugai}},\
  and\ \bibinfo {author} {\bibfnamefont {H.}~\bibnamefont {Suzuki}},\ }\href
  {https://doi.org/10.1143/JPSJ.74.1674} {\bibfield  {journal} {\bibinfo
  {journal} {Journal of the Physical Society of Japan}\ }\textbf {\bibinfo
  {volume} {74}},\ \bibinfo {pages} {1674} (\bibinfo {year}
  {2005})}\BibitemShut {NoStop}%
\bibitem [{\citenamefont {Kane}\ and\ \citenamefont
  {Mele}(2005{\natexlab{a}})}]{2005-PRL-KM-Z2}%
  \BibitemOpen
  \bibfield  {author} {\bibinfo {author} {\bibfnamefont {C.~L.}\ \bibnamefont
  {Kane}}\ and\ \bibinfo {author} {\bibfnamefont {E.~J.}\ \bibnamefont
  {Mele}},\ }\href {https://doi.org/10.1103/PhysRevLett.95.146802} {\bibfield
  {journal} {\bibinfo  {journal} {Phys. Rev. Lett.}\ }\textbf {\bibinfo
  {volume} {95}},\ \bibinfo {pages} {146802} (\bibinfo {year}
  {2005}{\natexlab{a}})}\BibitemShut {NoStop}%
\bibitem [{\citenamefont {Kane}\ and\ \citenamefont
  {Mele}(2005{\natexlab{b}})}]{2005-PRL-KM}%
  \BibitemOpen
  \bibfield  {author} {\bibinfo {author} {\bibfnamefont {C.~L.}\ \bibnamefont
  {Kane}}\ and\ \bibinfo {author} {\bibfnamefont {E.~J.}\ \bibnamefont
  {Mele}},\ }\href {https://doi.org/10.1103/PhysRevLett.95.226801} {\bibfield
  {journal} {\bibinfo  {journal} {Phys. Rev. Lett.}\ }\textbf {\bibinfo
  {volume} {95}},\ \bibinfo {pages} {226801} (\bibinfo {year}
  {2005}{\natexlab{b}})}\BibitemShut {NoStop}%
\bibitem [{\citenamefont {Xu}\ \emph {et~al.}(2019)\citenamefont {Xu},
  \citenamefont {Sheng},\ and\ \citenamefont {Yang}}]{YFYang-UTe}%
  \BibitemOpen
  \bibfield  {author} {\bibinfo {author} {\bibfnamefont {Y.}~\bibnamefont
  {Xu}}, \bibinfo {author} {\bibfnamefont {Y.}~\bibnamefont {Sheng}},\ and\
  \bibinfo {author} {\bibfnamefont {Y.-f.}\ \bibnamefont {Yang}},\ }\href
  {https://doi.org/10.1103/PhysRevLett.123.217002} {\bibfield  {journal}
  {\bibinfo  {journal} {Phys. Rev. Lett.}\ }\textbf {\bibinfo {volume} {123}},\
  \bibinfo {pages} {217002} (\bibinfo {year} {2019})}\BibitemShut {NoStop}%
\bibitem [{\citenamefont {Duan}\ \emph {et~al.}(2021)\citenamefont {Duan},
  \citenamefont {Baumbach}, \citenamefont {Podlesnyak}, \citenamefont {Deng},
  \citenamefont {Moir}, \citenamefont {Breindel}, \citenamefont {Maple},
  \citenamefont {Nica}, \citenamefont {Si},\ and\ \citenamefont
  {Dai}}]{QM-UTe2-nature}%
  \BibitemOpen
  \bibfield  {author} {\bibinfo {author} {\bibfnamefont {C.}~\bibnamefont
  {Duan}}, \bibinfo {author} {\bibfnamefont {R.~E.}\ \bibnamefont {Baumbach}},
  \bibinfo {author} {\bibfnamefont {A.}~\bibnamefont {Podlesnyak}}, \bibinfo
  {author} {\bibfnamefont {Y.}~\bibnamefont {Deng}}, \bibinfo {author}
  {\bibfnamefont {C.}~\bibnamefont {Moir}}, \bibinfo {author} {\bibfnamefont
  {A.~J.}\ \bibnamefont {Breindel}}, \bibinfo {author} {\bibfnamefont {M.~B.}\
  \bibnamefont {Maple}}, \bibinfo {author} {\bibfnamefont {E.~M.}\ \bibnamefont
  {Nica}}, \bibinfo {author} {\bibfnamefont {Q.}~\bibnamefont {Si}},\ and\
  \bibinfo {author} {\bibfnamefont {P.}~\bibnamefont {Dai}},\ }\href@noop {}
  {\bibfield  {journal} {\bibinfo  {journal} {Nature}\ }\textbf {\bibinfo
  {volume} {600}},\ \bibinfo {pages} {636} (\bibinfo {year}
  {2021})}\BibitemShut {NoStop}%
\bibitem [{\citenamefont {Miao}\ \emph {et~al.}(2020)\citenamefont {Miao},
  \citenamefont {Liu}, \citenamefont {Xu}, \citenamefont {Kotta}, \citenamefont
  {Kang}, \citenamefont {Ran}, \citenamefont {Paglione}, \citenamefont
  {Kotliar}, \citenamefont {Butch}, \citenamefont {Denlinger},\ and\
  \citenamefont {Wray}}]{Paglione-PRL-ARPES}%
  \BibitemOpen
  \bibfield  {author} {\bibinfo {author} {\bibfnamefont {L.}~\bibnamefont
  {Miao}}, \bibinfo {author} {\bibfnamefont {S.}~\bibnamefont {Liu}}, \bibinfo
  {author} {\bibfnamefont {Y.}~\bibnamefont {Xu}}, \bibinfo {author}
  {\bibfnamefont {E.~C.}\ \bibnamefont {Kotta}}, \bibinfo {author}
  {\bibfnamefont {C.-J.}\ \bibnamefont {Kang}}, \bibinfo {author}
  {\bibfnamefont {S.}~\bibnamefont {Ran}}, \bibinfo {author} {\bibfnamefont
  {J.}~\bibnamefont {Paglione}}, \bibinfo {author} {\bibfnamefont
  {G.}~\bibnamefont {Kotliar}}, \bibinfo {author} {\bibfnamefont {N.~P.}\
  \bibnamefont {Butch}}, \bibinfo {author} {\bibfnamefont {J.~D.}\ \bibnamefont
  {Denlinger}},\ and\ \bibinfo {author} {\bibfnamefont {L.~A.}\ \bibnamefont
  {Wray}},\ }\href {https://doi.org/10.1103/PhysRevLett.124.076401} {\bibfield
  {journal} {\bibinfo  {journal} {Phys. Rev. Lett.}\ }\textbf {\bibinfo
  {volume} {124}},\ \bibinfo {pages} {076401} (\bibinfo {year}
  {2020})}\BibitemShut {NoStop}%
\bibitem [{\citenamefont {Eo}\ \emph {et~al.}(2022)\citenamefont {Eo},
  \citenamefont {Liu}, \citenamefont {Saha}, \citenamefont {Kim}, \citenamefont
  {Ran}, \citenamefont {Horn}, \citenamefont {Hodovanets}, \citenamefont
  {Collini}, \citenamefont {Metz}, \citenamefont {Fuhrman}, \citenamefont
  {Nevidomskyy}, \citenamefont {Denlinger}, \citenamefont {Butch},
  \citenamefont {Fuhrer}, \citenamefont {Wray},\ and\ \citenamefont
  {Paglione}}]{rho-c-axis-PRB}%
  \BibitemOpen
  \bibfield  {author} {\bibinfo {author} {\bibfnamefont {Y.~S.}\ \bibnamefont
  {Eo}}, \bibinfo {author} {\bibfnamefont {S.}~\bibnamefont {Liu}}, \bibinfo
  {author} {\bibfnamefont {S.~R.}\ \bibnamefont {Saha}}, \bibinfo {author}
  {\bibfnamefont {H.}~\bibnamefont {Kim}}, \bibinfo {author} {\bibfnamefont
  {S.}~\bibnamefont {Ran}}, \bibinfo {author} {\bibfnamefont {J.~A.}\
  \bibnamefont {Horn}}, \bibinfo {author} {\bibfnamefont {H.}~\bibnamefont
  {Hodovanets}}, \bibinfo {author} {\bibfnamefont {J.}~\bibnamefont {Collini}},
  \bibinfo {author} {\bibfnamefont {T.}~\bibnamefont {Metz}}, \bibinfo {author}
  {\bibfnamefont {W.~T.}\ \bibnamefont {Fuhrman}}, \bibinfo {author}
  {\bibfnamefont {A.~H.}\ \bibnamefont {Nevidomskyy}}, \bibinfo {author}
  {\bibfnamefont {J.~D.}\ \bibnamefont {Denlinger}}, \bibinfo {author}
  {\bibfnamefont {N.~P.}\ \bibnamefont {Butch}}, \bibinfo {author}
  {\bibfnamefont {M.~S.}\ \bibnamefont {Fuhrer}}, \bibinfo {author}
  {\bibfnamefont {L.~A.}\ \bibnamefont {Wray}},\ and\ \bibinfo {author}
  {\bibfnamefont {J.}~\bibnamefont {Paglione}},\ }\href
  {https://doi.org/10.1103/PhysRevB.106.L060505} {\bibfield  {journal}
  {\bibinfo  {journal} {Phys. Rev. B}\ }\textbf {\bibinfo {volume} {106}},\
  \bibinfo {pages} {L060505} (\bibinfo {year} {2022})}\BibitemShut {NoStop}%
\bibitem [{\citenamefont {Braithwaite}\ \emph {et~al.}(2019)\citenamefont
  {Braithwaite}, \citenamefont {Vali{\v s}ka}, \citenamefont {Knebel},
  \citenamefont {Lapertot}, \citenamefont {Brison}, \citenamefont {Pourret},
  \citenamefont {Zhitomirsky}, \citenamefont {Flouquet}, \citenamefont
  {Honda},\ and\ \citenamefont {Aoki}}]{Braithwaite-CommPhys-PBphaseDiag}%
  \BibitemOpen
  \bibfield  {author} {\bibinfo {author} {\bibfnamefont {D.}~\bibnamefont
  {Braithwaite}}, \bibinfo {author} {\bibfnamefont {M.}~\bibnamefont {Vali{\v
  s}ka}}, \bibinfo {author} {\bibfnamefont {G.}~\bibnamefont {Knebel}},
  \bibinfo {author} {\bibfnamefont {G.}~\bibnamefont {Lapertot}}, \bibinfo
  {author} {\bibfnamefont {J.~P.}\ \bibnamefont {Brison}}, \bibinfo {author}
  {\bibfnamefont {A.}~\bibnamefont {Pourret}}, \bibinfo {author} {\bibfnamefont
  {M.~E.}\ \bibnamefont {Zhitomirsky}}, \bibinfo {author} {\bibfnamefont
  {J.}~\bibnamefont {Flouquet}}, \bibinfo {author} {\bibfnamefont
  {F.}~\bibnamefont {Honda}},\ and\ \bibinfo {author} {\bibfnamefont
  {D.}~\bibnamefont {Aoki}},\ }\href
  {https://doi.org/10.1038/s42005-019-0248-z} {\bibfield  {journal} {\bibinfo
  {journal} {Communications Physics}\ }\textbf {\bibinfo {volume} {2}},\
  \bibinfo {pages} {147} (\bibinfo {year} {2019})}\BibitemShut {NoStop}%
\bibitem [{\citenamefont {Zhao}\ \emph {et~al.}(2019)\citenamefont {Zhao},
  \citenamefont {Zhang}, \citenamefont {Lyu}, \citenamefont {Bachus},
  \citenamefont {Tokiwa}, \citenamefont {Gegenwart}, \citenamefont {Zhang},
  \citenamefont {Cheng}, \citenamefont {Yang}, \citenamefont {Chen},
  \citenamefont {Isikawa}, \citenamefont {Si}, \citenamefont {Steglich},\ and\
  \citenamefont {Sun}}]{2019-Sun-CePdAl}%
  \BibitemOpen
  \bibfield  {author} {\bibinfo {author} {\bibfnamefont {H.}~\bibnamefont
  {Zhao}}, \bibinfo {author} {\bibfnamefont {J.}~\bibnamefont {Zhang}},
  \bibinfo {author} {\bibfnamefont {M.}~\bibnamefont {Lyu}}, \bibinfo {author}
  {\bibfnamefont {S.}~\bibnamefont {Bachus}}, \bibinfo {author} {\bibfnamefont
  {Y.}~\bibnamefont {Tokiwa}}, \bibinfo {author} {\bibfnamefont
  {P.}~\bibnamefont {Gegenwart}}, \bibinfo {author} {\bibfnamefont
  {S.}~\bibnamefont {Zhang}}, \bibinfo {author} {\bibfnamefont
  {J.}~\bibnamefont {Cheng}}, \bibinfo {author} {\bibfnamefont {Y.-f.}\
  \bibnamefont {Yang}}, \bibinfo {author} {\bibfnamefont {G.}~\bibnamefont
  {Chen}}, \bibinfo {author} {\bibfnamefont {Y.}~\bibnamefont {Isikawa}},
  \bibinfo {author} {\bibfnamefont {Q.}~\bibnamefont {Si}}, \bibinfo {author}
  {\bibfnamefont {F.}~\bibnamefont {Steglich}},\ and\ \bibinfo {author}
  {\bibfnamefont {P.}~\bibnamefont {Sun}},\ }\href
  {https://doi.org/10.1038/s41567-019-0666-6} {\bibfield  {journal} {\bibinfo
  {journal} {Nat. Phys.}\ }\textbf {\bibinfo {volume} {15}},\ \bibinfo {pages}
  {1261} (\bibinfo {year} {2019})}\BibitemShut {NoStop}%
\bibitem [{\citenamefont {Aoki}\ \emph {et~al.}()\citenamefont {Aoki},
  \citenamefont {Nakamura}, \citenamefont {Honda}, \citenamefont {Li},
  \citenamefont {Homma}, \citenamefont {Shimizu}, \citenamefont {Sato},
  \citenamefont {Knebel}, \citenamefont {Brison}, \citenamefont {Pourret},
  \citenamefont {Braithwaite}, \citenamefont {Lapertot}, \citenamefont {Niu},
  \citenamefont {Vališka}, \citenamefont {Harima},\ and\ \citenamefont
  {Flouquet}}]{Aoki-JPSJ-2020-FM}%
  \BibitemOpen
  \bibfield  {author} {\bibinfo {author} {\bibfnamefont {D.}~\bibnamefont
  {Aoki}}, \bibinfo {author} {\bibfnamefont {A.}~\bibnamefont {Nakamura}},
  \bibinfo {author} {\bibfnamefont {F.}~\bibnamefont {Honda}}, \bibinfo
  {author} {\bibfnamefont {D.}~\bibnamefont {Li}}, \bibinfo {author}
  {\bibfnamefont {Y.}~\bibnamefont {Homma}}, \bibinfo {author} {\bibfnamefont
  {Y.}~\bibnamefont {Shimizu}}, \bibinfo {author} {\bibfnamefont {Y.~J.}\
  \bibnamefont {Sato}}, \bibinfo {author} {\bibfnamefont {G.}~\bibnamefont
  {Knebel}}, \bibinfo {author} {\bibfnamefont {J.-P.}\ \bibnamefont {Brison}},
  \bibinfo {author} {\bibfnamefont {A.}~\bibnamefont {Pourret}}, \bibinfo
  {author} {\bibfnamefont {D.}~\bibnamefont {Braithwaite}}, \bibinfo {author}
  {\bibfnamefont {G.}~\bibnamefont {Lapertot}}, \bibinfo {author}
  {\bibfnamefont {Q.}~\bibnamefont {Niu}}, \bibinfo {author} {\bibfnamefont
  {M.}~\bibnamefont {Vališka}}, \bibinfo {author} {\bibfnamefont
  {H.}~\bibnamefont {Harima}},\ and\ \bibinfo {author} {\bibfnamefont
  {J.}~\bibnamefont {Flouquet}},\ }\bibinfo {title} {{Spin-Triplet
  Superconductivity in UTe$_2$ and Ferromagnetic Superconductors}},\ in\ \href
  {https://doi.org/10.7566/JPSCP.30.011065} {\emph {\bibinfo {booktitle}
  {Proceedings of the International Conference on Strongly Correlated Electron
  Systems (SCES2019)}}},\ \Eprint
  {https://arxiv.org/abs/https://journals.jps.jp/doi/pdf/10.7566/JPSCP.30.011065}
  {https://journals.jps.jp/doi/pdf/10.7566/JPSCP.30.011065} \BibitemShut
  {NoStop}%
\bibitem [{\citenamefont {Metz}\ \emph {et~al.}(2019)\citenamefont {Metz},
  \citenamefont {Bae}, \citenamefont {Ran}, \citenamefont {Liu}, \citenamefont
  {Eo}, \citenamefont {Fuhrman}, \citenamefont {Agterberg}, \citenamefont
  {Anlage}, \citenamefont {Butch},\ and\ \citenamefont
  {Paglione}}]{point-node-Paglione}%
  \BibitemOpen
  \bibfield  {author} {\bibinfo {author} {\bibfnamefont {T.}~\bibnamefont
  {Metz}}, \bibinfo {author} {\bibfnamefont {S.}~\bibnamefont {Bae}}, \bibinfo
  {author} {\bibfnamefont {S.}~\bibnamefont {Ran}}, \bibinfo {author}
  {\bibfnamefont {I.-L.}\ \bibnamefont {Liu}}, \bibinfo {author} {\bibfnamefont
  {Y.~S.}\ \bibnamefont {Eo}}, \bibinfo {author} {\bibfnamefont {W.~T.}\
  \bibnamefont {Fuhrman}}, \bibinfo {author} {\bibfnamefont {D.~F.}\
  \bibnamefont {Agterberg}}, \bibinfo {author} {\bibfnamefont {S.~M.}\
  \bibnamefont {Anlage}}, \bibinfo {author} {\bibfnamefont {N.~P.}\
  \bibnamefont {Butch}},\ and\ \bibinfo {author} {\bibfnamefont
  {J.}~\bibnamefont {Paglione}},\ }\href
  {https://doi.org/10.1103/PhysRevB.100.220504} {\bibfield  {journal} {\bibinfo
   {journal} {Phys. Rev. B}\ }\textbf {\bibinfo {volume} {100}},\ \bibinfo
  {pages} {220504} (\bibinfo {year} {2019})}\BibitemShut {NoStop}%
\bibitem [{\citenamefont {Burdin}\ \emph {et~al.}(2000)\citenamefont {Burdin},
  \citenamefont {Georges},\ and\ \citenamefont {Grempel}}]{Burdin-PRL}%
  \BibitemOpen
  \bibfield  {author} {\bibinfo {author} {\bibfnamefont {S.}~\bibnamefont
  {Burdin}}, \bibinfo {author} {\bibfnamefont {A.}~\bibnamefont {Georges}},\
  and\ \bibinfo {author} {\bibfnamefont {D.~R.}\ \bibnamefont {Grempel}},\
  }\href {https://doi.org/10.1103/PhysRevLett.85.1048} {\bibfield  {journal}
  {\bibinfo  {journal} {Phys. Rev. Lett.}\ }\textbf {\bibinfo {volume} {85}},\
  \bibinfo {pages} {1048} (\bibinfo {year} {2000})}\BibitemShut {NoStop}%
\bibitem [{\citenamefont {Hayes}\ \emph {et~al.}(2021)\citenamefont {Hayes},
  \citenamefont {Wei}, \citenamefont {Metz}, \citenamefont {Zhang},
  \citenamefont {Eo}, \citenamefont {Ran}, \citenamefont {Saha}, \citenamefont
  {Collini}, \citenamefont {Butch}, \citenamefont {Agterberg}, \citenamefont
  {Kapitulnik},\ and\ \citenamefont {Paglione}}]{Hayes-KerrEff-Science}%
  \BibitemOpen
  \bibfield  {author} {\bibinfo {author} {\bibfnamefont {I.~M.}\ \bibnamefont
  {Hayes}}, \bibinfo {author} {\bibfnamefont {D.~S.}\ \bibnamefont {Wei}},
  \bibinfo {author} {\bibfnamefont {T.}~\bibnamefont {Metz}}, \bibinfo {author}
  {\bibfnamefont {J.}~\bibnamefont {Zhang}}, \bibinfo {author} {\bibfnamefont
  {Y.~S.}\ \bibnamefont {Eo}}, \bibinfo {author} {\bibfnamefont
  {S.}~\bibnamefont {Ran}}, \bibinfo {author} {\bibfnamefont {S.~R.}\
  \bibnamefont {Saha}}, \bibinfo {author} {\bibfnamefont {J.}~\bibnamefont
  {Collini}}, \bibinfo {author} {\bibfnamefont {N.~P.}\ \bibnamefont {Butch}},
  \bibinfo {author} {\bibfnamefont {D.~F.}\ \bibnamefont {Agterberg}}, \bibinfo
  {author} {\bibfnamefont {A.}~\bibnamefont {Kapitulnik}},\ and\ \bibinfo
  {author} {\bibfnamefont {J.}~\bibnamefont {Paglione}},\ }\href
  {https://doi.org/10.1126/science.abb0272} {\bibfield  {journal} {\bibinfo
  {journal} {Science}\ }\textbf {\bibinfo {volume} {373}},\ \bibinfo {pages}
  {797} (\bibinfo {year} {2021})},\ \Eprint
  {https://arxiv.org/abs/https://www.science.org/doi/pdf/10.1126/science.abb0272}
  {https://www.science.org/doi/pdf/10.1126/science.abb0272} \BibitemShut
  {NoStop}%
\bibitem [{\citenamefont {Shishidou}\ \emph {et~al.}(2021)\citenamefont
  {Shishidou}, \citenamefont {Suh}, \citenamefont {Brydon}, \citenamefont
  {Weinert},\ and\ \citenamefont {Agterberg}}]{Agterberg-topo-band-PRB}%
  \BibitemOpen
  \bibfield  {author} {\bibinfo {author} {\bibfnamefont {T.}~\bibnamefont
  {Shishidou}}, \bibinfo {author} {\bibfnamefont {H.~G.}\ \bibnamefont {Suh}},
  \bibinfo {author} {\bibfnamefont {P.~M.~R.}\ \bibnamefont {Brydon}}, \bibinfo
  {author} {\bibfnamefont {M.}~\bibnamefont {Weinert}},\ and\ \bibinfo {author}
  {\bibfnamefont {D.~F.}\ \bibnamefont {Agterberg}},\ }\href
  {https://doi.org/10.1103/PhysRevB.103.104504} {\bibfield  {journal} {\bibinfo
   {journal} {Phys. Rev. B}\ }\textbf {\bibinfo {volume} {103}},\ \bibinfo
  {pages} {104504} (\bibinfo {year} {2021})}\BibitemShut {NoStop}%
\bibitem [{\citenamefont {{Hazra}}\ and\ \citenamefont
  {{Coleman}}(2022)}]{Coleman-2022-UTe}%
  \BibitemOpen
  \bibfield  {author} {\bibinfo {author} {\bibfnamefont {T.}~\bibnamefont
  {{Hazra}}}\ and\ \bibinfo {author} {\bibfnamefont {P.}~\bibnamefont
  {{Coleman}}},\ }\href@noop {} {\bibinfo {title} {{Triplet pairing mechanisms
  from Hund's-Kondo models: applications to UTe$_{2}$ and CeRh$_{2}$As$_{2}$}}}
  (\bibinfo {year} {2022}),\ \Eprint {https://arxiv.org/abs/2205.13529}
  {arXiv:2205.13529 [cond-mat.supr-con]} \BibitemShut {NoStop}%
\bibitem [{\citenamefont {Rosa}\ \emph {et~al.}(2022)\citenamefont {Rosa},
  \citenamefont {Weiland}, \citenamefont {Fender}, \citenamefont {Scott},
  \citenamefont {Ronning}, \citenamefont {Thompson}, \citenamefont {Bauer},\
  and\ \citenamefont {Thomas}}]{Rosa-SingleTran-CommMat}%
  \BibitemOpen
  \bibfield  {author} {\bibinfo {author} {\bibfnamefont {P.~F.~S.}\
  \bibnamefont {Rosa}}, \bibinfo {author} {\bibfnamefont {A.}~\bibnamefont
  {Weiland}}, \bibinfo {author} {\bibfnamefont {S.~S.}\ \bibnamefont {Fender}},
  \bibinfo {author} {\bibfnamefont {B.~L.}\ \bibnamefont {Scott}}, \bibinfo
  {author} {\bibfnamefont {F.}~\bibnamefont {Ronning}}, \bibinfo {author}
  {\bibfnamefont {J.~D.}\ \bibnamefont {Thompson}}, \bibinfo {author}
  {\bibfnamefont {E.~D.}\ \bibnamefont {Bauer}},\ and\ \bibinfo {author}
  {\bibfnamefont {S.~M.}\ \bibnamefont {Thomas}},\ }\href
  {https://doi.org/10.1038/s43246-022-00254-2} {\bibfield  {journal} {\bibinfo
  {journal} {Communications Materials}\ }\textbf {\bibinfo {volume} {3}},\
  \bibinfo {pages} {33} (\bibinfo {year} {2022})}\BibitemShut {NoStop}%
\bibitem [{\citenamefont {Rosuel}\ \emph {et~al.}(2022)\citenamefont {Rosuel},
  \citenamefont {Marcenat}, \citenamefont {Knebel}, \citenamefont {Klein},
  \citenamefont {Pourret}, \citenamefont {Marquardt}, \citenamefont {Niu},
  \citenamefont {Rousseau}, \citenamefont {Demuer}, \citenamefont {Seyfarth},
  \citenamefont {Lapertot}, \citenamefont {Aoki}, \citenamefont {Braithwaite},
  \citenamefont {Flouquet},\ and\ \citenamefont
  {Brison}}]{Rousel-arxiv-singletransition}%
  \BibitemOpen
  \bibfield  {author} {\bibinfo {author} {\bibfnamefont {A.}~\bibnamefont
  {Rosuel}}, \bibinfo {author} {\bibfnamefont {C.}~\bibnamefont {Marcenat}},
  \bibinfo {author} {\bibfnamefont {G.}~\bibnamefont {Knebel}}, \bibinfo
  {author} {\bibfnamefont {T.}~\bibnamefont {Klein}}, \bibinfo {author}
  {\bibfnamefont {A.}~\bibnamefont {Pourret}}, \bibinfo {author} {\bibfnamefont
  {N.}~\bibnamefont {Marquardt}}, \bibinfo {author} {\bibfnamefont
  {Q.}~\bibnamefont {Niu}}, \bibinfo {author} {\bibfnamefont {S.}~\bibnamefont
  {Rousseau}}, \bibinfo {author} {\bibfnamefont {A.}~\bibnamefont {Demuer}},
  \bibinfo {author} {\bibfnamefont {G.}~\bibnamefont {Seyfarth}}, \bibinfo
  {author} {\bibfnamefont {G.}~\bibnamefont {Lapertot}}, \bibinfo {author}
  {\bibfnamefont {D.}~\bibnamefont {Aoki}}, \bibinfo {author} {\bibfnamefont
  {D.}~\bibnamefont {Braithwaite}}, \bibinfo {author} {\bibfnamefont
  {J.}~\bibnamefont {Flouquet}},\ and\ \bibinfo {author} {\bibfnamefont
  {J.-P.}\ \bibnamefont {Brison}},\ }\href
  {https://doi.org/10.48550/ARXIV.2205.04524} {\bibinfo {title} {Field-induced
  tuning of the pairing state in a superconductor}} (\bibinfo {year}
  {2022})\BibitemShut {NoStop}%
\bibitem [{\citenamefont {Sato}\ and\ \citenamefont
  {Fujimoto}(2009)}]{sato-PRB-NSC}%
  \BibitemOpen
  \bibfield  {author} {\bibinfo {author} {\bibfnamefont {M.}~\bibnamefont
  {Sato}}\ and\ \bibinfo {author} {\bibfnamefont {S.}~\bibnamefont
  {Fujimoto}},\ }\href {https://doi.org/10.1103/PhysRevB.79.094504} {\bibfield
  {journal} {\bibinfo  {journal} {Phys. Rev. B}\ }\textbf {\bibinfo {volume}
  {79}},\ \bibinfo {pages} {094504} (\bibinfo {year} {2009})}\BibitemShut
  {NoStop}%
\end{thebibliography}
\end{document}